\documentclass{aa}
\usepackage{graphicx,amssymb}

\newcommand{\degrees}{\ensuremath{^\circ}}

\begin{document}
   \title{Beam mismatch effects in Cosmic Microwave Background 
     polarization measurements}

   \titlerunning{Beam mismatch in CMB polarization measurements}

   \author{
          C. Rosset\inst{1,4}
          \and
          V. B. Yurchenko\inst{2}
	  \and
          J. Delabrouille\inst{1}
          \and
          J. Kaplan\inst{1}
	  \and
          Y. Giraud--H\'eraud\inst{1}
          \and
          J.--M. Lamarre\inst{3}
          \and
          J.~A. Murphy\inst{2}
          }

   \institute{
              APC, Universit\'e Paris 7, CNRS/IN2P3, 10 rue Alice Domon 
et L\'eonie Duquet 75205 Paris Cedex 13, France\\
             \email{delabrouille@apc.univ-paris7.fr;
	       kaplan@apc.univ-paris7.fr, ygh@apc.univ-paris7.fr}
         \and
             Experimental Physics Department, National University 
of Ireland, Maynooth, Co. Kildare, Ireland\\
              \email{amurphy@may.ie; v.yurchenko@may.ie}
         \and
             LERMA, Observatoire de Paris, 61 Av. de l'Observatoire, 
75014, Paris, France\\
              \email{jean-michel.lamarre@obspm.fr}
         \and 
	     LAL, Universit\'e Paris-Sud 11, CNRS/IN2P3, B.P. 34, 91898 ORSAY Cedex, France\\
              \email{rosset@lal.in2p3.fr}
             }

   \date{Received November $1^{\mathrm{st}}$, 2004; accepted December, $6^{\mathrm{th}}$, 2006}

   \abstract{Measurement of cosmic microwave background polarization
   is today a major goal of observational cosmology. The level of the
   signal to measure, however, makes it very sensitive to various
   systematic effects. In the case of Planck, which measures
   polarization by combining data from various detectors, the beam
   asymmetry can induce a conversion of temperature signals to
   polarization signals or a polarization mode mixing. In this paper,
   we investigate this effect using realistic simulated beams and
   propose a first-order method to correct the polarization power
   spectra for the induced systematic effect.
     
   }

   \maketitle
%

\section{Introduction}
After the success of \emph{COBE} (Smoot et al, \cite{smoot}) and
\emph{WMAP} (Bennett {\em et al.} \cite{Bennett} , the Planck mission,
to be launched by ESA in early 2007, is the third generation space
mission dedicated to the measurement of the properties of the Cosmic
Microwave Background (CMB). About 20 times more sensitive than
\emph{WMAP}, Planck will observe the full sky in the millimeter and
sub-millimeter domain in nine frequency channels centered around
frequencies ranging from 30 to 70 Ghz (for the Low Frequncy
Instrument, or LFI) and from 100 to 850 GHz (for the High Frequency
Instrument or HFI). Of these channels, the seven at lowest frequencies
--- from 30 to 350 GHz, are polarization sensitive.

Temperature anisotropies have been detected by many experiments now,
the most recent of which detect a series of acoustic peaks in the CMB
spatial power spectrum (de Bernardis et al, \cite{boom1}, Hanany et
al, \cite{maxima1}, Beno\^{i}t et al, \cite{archeops1}, Hinshaw et al,
\cite{wmapcl}), confirm the Gaussianity of observable CMB fluctuations
(Komatsu et al, \cite{wmapgauss}, though wavelet methods have detected
presence of non-Gaussianity in \emph{WMAP} data, Vielva et al,
\cite{vielva}), and demonstrate the spatial flatness of the Universe
(Netterfield et al, \cite{Netterfield}, Lee et al, \cite{maxima2},
Beno\^{i}t et al, \cite{archeops2}, Spergel et al, \cite{wmapcp},
Spergel et al, \cite{wmapcp3}). This provides compelling evidence that
the primordial perturbations indeed have been generated during an
inflationary period in the very early Universe. The next challenge now
is the precise measurement of polarization anisotropies and, in
particular, the detection of the pseudo-scalar part of the
polarization field (the $B$ modes of CMB polarization) which are
expected to carry the unambiguous signature of the energy scale of
inflation and of the potential of the inflationary field. The Planck
mission will be the first experiment able to constrain significantly
these $B$ modes over the full sky and hence to measure them on very
large scales.

The first detection of CMB polarization at one degree angular scale of
resolution, at a level compatible with predictions of the standard
cosmological scenario, has been announced by Kovac {\em et al}
(\cite{Kovac}) (see also Leitch {\em et al} \cite{Leitch}). Since
then, CBI and CAPMAP have also obtained significant detection of CMB
E--mode polarization (Readhead {\em et al.}  \cite{Readhead} and
Barkats {\em et al.} \cite{Barkats}). More recently, the {\emph
Boomerang} (Piacentini {\it et al} \cite{Piacentini}, Montroy {\it et
al} \cite{Montroy}) and \emph{WMAP} (Kogut {\em et al.} \cite{Kogut},
Page {\it et al.}  \cite{Page}) teams have obtained a measurement of
the temperature-polarization correlation and E--mode spectrum
compatible with cosmological model. No significant constraint on
B--mode at degree angular scales exist today.

While the measurement of the temperature and polarization auto and
cross power spectra of the CMB carries a wealth of information about
cosmological parameters and about scenarios for the generation of the
seeds for structure formation, some near--degeneracies exist which
require extremely precise measurements. In particular, a very precise
control of systematic errors is required to constrain parameters which
impact these anisotropies and polarization fields at a very low level.

Many sources of systematic errors are potentially a problem for
polarization measurements. In particular, the shape of the beams of
the instrument need to be known with extreme precision. In addition, when
the measurements of several detectors are combined to obtain
polarization signals, it is required that the responses of these
detectors be matched precisely in terms of cross--calibration, 
beam shape, spectral response, etc.

Measurements of Planck telescope beams in the actual operation
conditions are not to be made on ground. Also, there are no polarized
astrophysical sources for the in-flight beam calibration. Therefore,
one should rely on numerical simulations of the beams and
self-correcting algorithms of data processing that should allow
efficient elimination of systematic errors. In polarization
measurements, a significant systematic error would arise due to
elliptical shapes of telescope beams which appear, mainly, due to
ellipsoidal shape of telescope mirrors introducing astigmatic
aberrations and other beam imperfections even with the otherwise ideal
mirror surfaces.

In this paper, we investigate the impact of beam imperfections on the
measurement of polarization power spectra. We then discuss a method
for first-order correction of the effect of these imperfections. To
illustrate this method, we apply it to the case of Planck HFI
polarization measurement.

The remainder of this paper is organized as follows. In section
\ref{sec:problem}, we discuss the issue of beam shape mismatch for the
detection of CMB polarization. Section \ref{sec:readouts} is dedicated to the
computation of simulated readouts using the realistic beams
described in Section~\ref{sec:problem} and the reconstruction of
polarized power spectra. In Section \ref{sec:correction}, we present a
method to correct for the systematic bias in the $B$ mode power
spectrum induced by the asymmetry of the beams. Finally, Section
\ref{sec:conclusions} draws the conclusions.


\section{The beam--mismatch problem in polarization measurements}
\label{sec:problem}

The HFI polarimeters employ Polarization Sensitive Bolometers (PSB)
cooled down to the temperature of 100 mK by a space $^3$He~--$^4$He
dilution fridge. These devices, also used on Boomerang (though at
300~mK), are presently the most sensitive operational detectors for
CMB polarization measurements (Jones~{\it et al.}, \cite{psb},
Montroy~{\it et al.}, \cite{b2k}). Each PSB measures the power of the
CMB field component along one linear direction specified by the PSB
orientation (Turner et al, \cite{PSB}).

Ideally, the PSBs are combined in pairs,
each pair placed at the rear side of respective HFI horn, with the two
PSB of the pair, $a$ and $b$, being oriented at $90^\circ$ of relative
angle and receiving the radiation from the same point on the sky. The
ideal polarimeters produce the measured signals (readouts):
\begin{equation}
\label{eqn:sa}
s_a = \frac{1}{2}(I + Q \cos 2\alpha + U \sin 2\alpha)
\end{equation}
\begin{equation}
\label{eqn:sb}
s_b = \frac{1}{2}(I - Q \cos 2\alpha - U \sin 2\alpha)
\end{equation}
where $I$, $Q$, $U$ are the Stokes parameters of incoming radiation
(see {\em e.g.} (Born and Wolf, \cite{born}) for the definition of
Stokes parameters) and $\alpha$ is the angle between the orientation
of the first ($a$) of two PSBs and the first ($x$) of two
orthogonal axes of the frame chosen for the representation of Stokes
parameters. The $V$ Stokes parameter does not enter Eqs.~(\ref{eqn:sa}),
(\ref{eqn:sb}) since the PSBs are designed to be, ideally, insensitive
to $V$ and, besides, $V$ is extremely small for the CMB radiation.

In practice, the detectors produce beam--integrated signals so that
equation~\ref{eqn:sa} is modified to (Kraus \cite{Kraus})
\begin{equation}
\label{eqn:sigconva}
s_a = \frac{1}{2}\int_{\rm beam} {\rm d}\Omega \int_{\rm band} {\rm
d}\nu \,({\widetilde I_a}I + {\widetilde Q_a}Q + {\widetilde U_a}U 
 + {\widetilde V_a}V )
\end{equation}
and similarly for $s_b$ where the PSB responses
${\widetilde I_a}(\vec x,\nu)$ etc are the telescope beam patterns of
Stokes parameters computed in transmitting mode and normalized to
unity at maximum, functions
of both the radiation frequency $\nu$ and the observation point $\vec
x$ (the $V$ term is neglected in most of the following discussion).
The responses of different polarimeters (a and b) should be adjusted
as much as possible (both in frequency and in angular pattern on the
sky) so that, ideally, one should have
\begin{eqnarray}
{\widetilde I_a}&=&{\widetilde I_b}\\
{\widetilde Q_a}&=& \ \ {\widetilde I_a} \cos 2\alpha, \quad
{\widetilde U_a}= \ \ {\widetilde I_a} \sin 2\alpha, \label{eqn:quia}\\
{\widetilde Q_b}&=&-{\widetilde I_b}\cos 2\alpha, \quad
{\widetilde U_b}=-{\widetilde I_b}\sin 2\alpha \label{eqn:quib}\\
{\widetilde V_a}&=&{\widetilde V_b}=0
\end{eqnarray}
where $\alpha$, similarly to the definition above, is the angle of
nominal orientation of polarimeter $a$ on the sky with respect to the
reference axis chosen for the definition of $Q$ and $U$. In this case,
with $\widetilde{I}_a = \widetilde{I}_b$, Eq.~(\ref{eqn:sigconva}) is
reduced to the form similar to Eqs.~(\ref{eqn:sa}) and~(\ref{eqn:sb}).

For simplicity, we approximate the PSB response as averaged over 
the frequency band of the particular channel, thus introducing 
the band--averaged beam patterns defined as
\begin{equation}
\label{eqn:response}
\widetilde I(\vec x) = \int_{\rm band} {\rm d}\nu \, 
{\widetilde I}(\vec x,\nu)
\end{equation}
and similarly for $\widetilde Q$ and $\widetilde U$ (for radiation
independent on $\nu$ on the beam width scale, this generates an exact
readout).

Ideally, the beam patterns on the sky ${\widetilde I}(\vec x)$ should
be as close as possible to a perfect Gaussian. Unfortunately, design
and construction imperfections, telescope aberrations, and optical misalignment
all generate small differences in the beam patterns, the impact of
which must be investigated accurately, especially for very sensitive
CMB polarization measurements.

\begin{figure}[tb]
\begin{center}
\includegraphics[width=8cm]{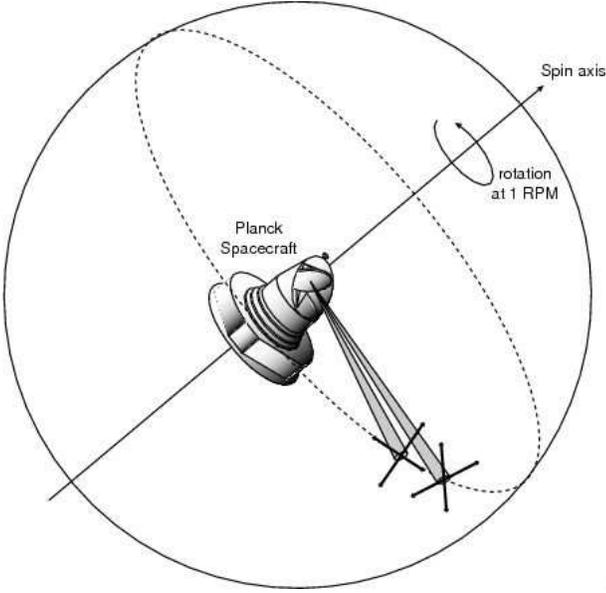}(a)
\includegraphics[width=8cm]{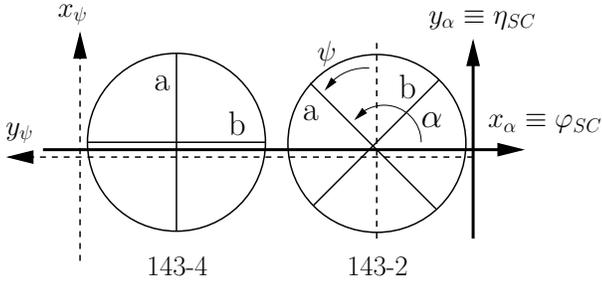}(b)
\caption{\label{fig:planck-pol-scanning} (a) The scanning of the
polarized detectors provides the measurements of intensity of the CMB
field components along four directions at each point on the scan
path. (b) Definition of axis specifying the Stokes parameters
reference frame as seen from the sky. The detector pair on the left
({\em e.g.} 143-4) measures the $Q$ Stokes parameter, while the pair
on the right ({\em e.g.} 143-2) measures $U$ when $Q$ and $U$ are
defined with respect to the $(x_\alpha,y_\alpha)$ frame (see
Appendix~\ref{sec:beams} for further detail on reference axis).
}
\label{fig:planck-scanning-QU}
\end{center}
\end{figure}

Measuring polarization, {\em i.e.} measuring the $I$, $Q$ and $U$
Stokes parameters, indeed involves combining several such measurements
with different angles $\alpha$ to separate the $I$, $Q$ and $U$
contributions. The Planck HFI detector set--up is such that the beams
of two horns with complementary pairs of PSB oriented at 45\degrees\
one pair with respect to the other follow each other on circular scan
paths on the sky as shown in Figs.~\ref{fig:planck-scanning-QU}
and~\ref{fig:focplane} (see Appendix~\ref{sec:beams} for further
detail and notations).  Then, in a system where reference
axes for defining $Q$ and $U$ are along the scan path ($x$) and
orthogonal to it ($y$), the four readouts $s_\alpha$ of this set of
PSB ($\alpha=0\degrees, 45\degrees, 90\degrees, 135\degrees$) allow
the direct measurement of $I$, $Q$ and $U$ as
\begin{eqnarray}
I &=& s_{0\degrees} + s_{90\degrees} = s_{45\degrees} + s_{135\degrees} \nonumber \\
\label{eqn:reciqu} Q &=& s_{0\degrees} - s_{90\degrees} \\
U & = & s_{45\degrees} - s_{135\degrees}. \nonumber
\end{eqnarray}

   \begin{figure}[tb]
   \centering
   \includegraphics[width=8cm]{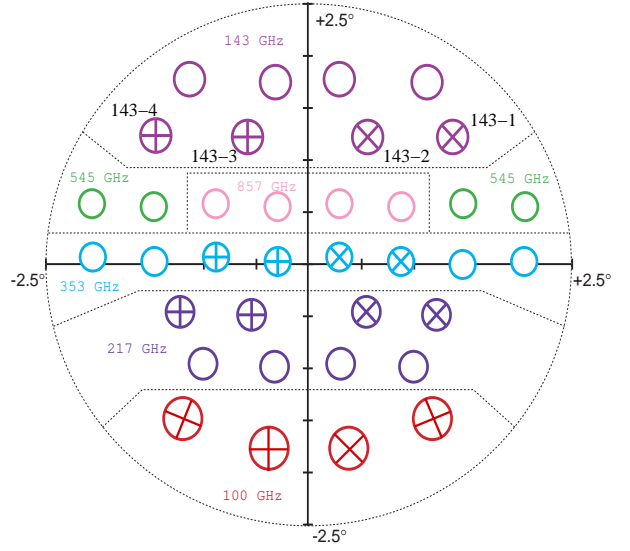}
   \caption{Planck focal plane unit (FPU) with polarization sensitive
   bolometers as seen from the sky. Complementary pairs of PSB
   detectors are arranged in two horns following each other while
   scanning the sky so that four detectors are in an optimized
   configuration for polarization measurement.}
         \label{fig:focplane}
   \end{figure}

In practice, when the responses ${\widetilde I_{0\degrees}}$ and
${\widetilde I_{90\degrees}}$ are not perfectly equal, there is a
small residual of $I$ in the estimate $\widehat Q$ of $Q$:
\begin{equation}
\label{eqn:residual}
{\widehat Q} = Q + \frac{1}{2}\int_{\rm beam} {\rm d}\Omega
({\widetilde I_{0\degrees}} - {\widetilde I_{90\degrees}})I.
\end{equation}
Similarly, there may be a small leakage of each Stokes component into
the others. These errors are a source of trouble for measuring $B$
mode especially, as they result in the leakages of $I$ into $E$ and
$B$ (possibly significant) and of $E$ into $B$ because $I \gg E \gg B$
on most scales.

This source of systematic effects for polarization measurements is not
specific to Planck. Any instrument measuring polarization in a similar
way, where signals proportional to $I$ need to be eliminated from the
measurements in order to obtain polarization data, may suffer
from this.

The quantitative investigation of the impact of such effects requires
a realistic estimate of mismatch between the companion beams, the
simulation of signal data using these beams, the reconstruction of
maps and of power spectra using these simulated data, and the
investigation of the correction of the effect by data processing
methods.

\begin{figure}[tb]
   \centering
   \includegraphics[width=8cm]{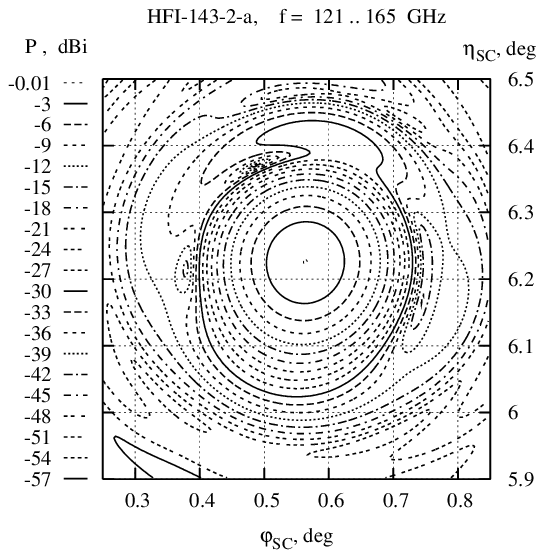}
(a)
   \includegraphics[width=8cm]{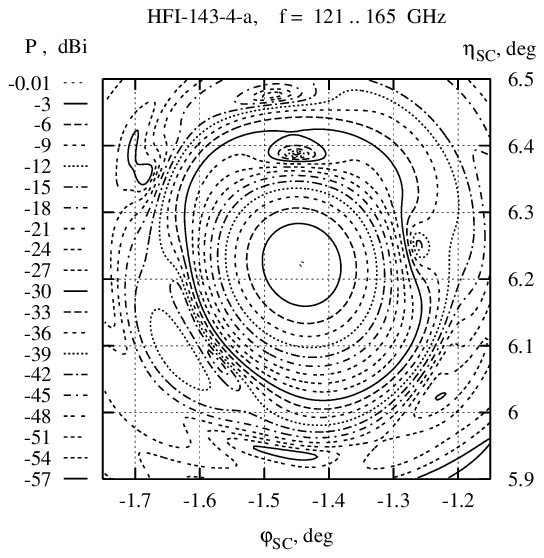}
(b)
      \caption{Broad--band power patterns ($\widetilde I$ responses) 
of the telescope beams to be superimposed on the sky for polarization 
measurements, (a) HFI-143-2a and (b) HFI-143-4a, in SC spherical frame 
on the sky, with the spin axis of telescope as a pole (isolevels are shown 
from the maximum down to $-60\,$dB with a step of $-3\,$dB).}
         \label{beams}
   \end{figure}

The computation of the Planck HFI beams is discussed in
Appendix~\ref{sec:beams} and in (Yurchenko {\it et al.},
\cite{Yurch2004}). Here, we will describe their main characteristics
relevant for the following sections. Because of telescope
aberrations, the shape of the intensity beams is
essentially Gaussian elliptic (down to nearly -30~dB) with the major
axis around 10\% longer than the minor axis (see
Fig.~\ref{beams}). Thanks to the use of PSB, the intensity beams of an
orthogonal pair of detectors within one horn are very close to each
other: the difference is at most 0.6\%. On the other hand, the
difference between beams of detectors in two different horns can be up
to 7\%: this difference is mainly due to the different orientation of
the beam ellipses. As emphasized in Appendix~\ref{sec:beams}, relaxing
the assumptions of perfect conductors and perfect alignment is not
expected to strongly modify the general shape of the beams. In the
next sections, we will refer to these computer-simulated beams as
``realistic beams''.

\section{Effect of beams on polarization power spectra}
\label{sec:readouts}

The main goal of this section is to study the systematic effect
induced on the power spectra estimation by realistic beams described
in the previous section, knowing that this effect will depend also on
the scanning strategy. As the Planck mission will scan the sky along
large opening angle circles, resulting in large parts of the sky where
the scans are mostly parallel, we have focused the study to the
observation of a $15^\circ \times 15^\circ$ region of the sky scanned
only along parallel directions. This restriction does not spoil the
interest of the study as the small scale distortion of the beams are
expected to affect mainly the small angular scales of the power
spectra. In addition, other experiments scanning only a fraction of
the sky are affected by the similar systematic effects. This
restriction also offers a practical advantage: the computation of the
effects of tiny beam mismatches on sub-beam scales requires a map
resolution better than the beam size. For the Planck HFI 143 GHz
channel, the resolution of about 7 arcminutes justifies models at
sub-arcminute scales. We have thus chosen to work on maps of
$2048\times 2048$ pixels of about 30 arcseconds each.

The following paragraphs describe the generation of CMB polarization
maps from power spectra, and the simulation of instrument signals.

\subsection{Generation of CMB polarization maps}

Simulated square maps of CMB intensity and polarization are generated
using the approximate relation between the power spectra in flat
($C(k)$) and spherical ($C_l$) coordinates: $k^2 C(k) \simeq
\left.l(l+1) C_l\right|_{l=k}$ (see, for example, White et al
\cite{White99}). The three maps of $T$, $E$ and $B$ are then computed
from three independent realizations of Gaussian white noise
$D^1(\mathbf{k})$, $D^2(\mathbf{k})$ and $D^3(\mathbf{k})$ as:
$$
a^{T,B}(\mathbf{k}) = D^{1,3}(\mathbf{k})\sqrt{C_l^{T,B}}
$$
and
$$
a^E(\mathbf{k}) =  D^1(\mathbf{k})
\frac{C_l^{TE}}{\sqrt{C_l^T}} + D^2(\mathbf{k})
\left(C_l^E-\frac{\left(C_l^{TE}\right)^2}{C_l^T}\right)^{1/2}
$$
so that the correlation between the $T$ and $E$ maps is taken into
account. $C_l$'s are the usual spectra describing the CMB temperature
and polarization. For our simulations, we used the cosmological
parameters from the \emph{WMAP} best fit model, except that we imposed
a tensor to scalar ratio of 0.1. The simulated maps include the
Gaussian part of the gravitational lensing effect of the $E$ mode.

\subsection{Simulation of instrument readouts}

The readouts must be computed from the $I$, $Q$ and $U$ Stokes
parameters. We thus need to convert the $E$ and $B$ maps to $Q$ and
$U$ using relations (38) in Zaldarriaga \& Seljak
\cite{Zaldarriaga97}:
\begin{equation}
\label{eqn:q_eb}
a^Q(\mathbf{k}) = a^E(\mathbf{k})\cos
  2\phi_\mathbf{k} - a^B(\mathbf{k}) \sin 2\phi_\mathbf{k}
\end{equation}
\begin{equation}
\label{eqn:u_eb}
a^U(\mathbf{k}) = a^E(\mathbf{k})\sin
  2\phi_\mathbf{k} + a^B(\mathbf{k}) \cos 2\phi_\mathbf{k}
\end{equation}
where $k_x+ik_y = k\mbox{e}^{i\phi_\mathbf{k}}$. The readout from one
detector is then obtained by convolving its $\widetilde{I}$,
$\widetilde{Q}$ and $\widetilde{U}$ beams with the $I$, $Q$ and $U$
maps from the sky and summing as in Eq.~(\ref{eqn:sigconva}). In the
case of the parallel scanning strategy we used, the convolution can be
easily done, once for all directions of observation, by multiplication
in Fourier space. Thus, we obtain four maps of readout signals, one
for each polarization channel, $s_{0\degrees}$, $s_{90\degrees}$,
$s_{45\degrees}$ and $s_{135\degrees}$, with polarization angles
$\alpha=0^\circ$, $90^\circ$, $45^\circ$ and $135^\circ$ with respect
to the $x$-axis of the map. With account of established focal plane
unit (FPU) notation of channels (see Appendix~\ref{sec:beams}), they
correspond, e.g., to the PSB channels 4b, 4a, 2b, and 2a,
respectively, of two horns HFI-143-4 and HFI-143-2 where $x$-axis is
the $\varphi_{SC}$-axis of spacecraft (SC, see
Appendix~\ref{sec:beams}) frame viewed from the sky
(Fig.~\ref{fig:planck-scanning-QU}b).

Since the goal of this work is to study only the systematic bias
induced on polarization power spectra, we do not add any white or
low-frequency noise to the signal, neither any other systematic
effects (Kaplan \& Delabrouille \cite{Kaplan}). These other
systematic effects will be studied in detail in a forthcoming
paper. In particular, we assume here that the time constant of
bolometers, which induce an elongation of the beams in the scanning
direction, has been corrected for.

\subsection{Reconstruction of the power spectra}
\label{sec:reconstruction}

The parallel scanning strategy allows us to reconstruct the $I$, $Q$
and $U$ maps from the readout maps using Eqs.~(\ref{eqn:reciqu}). The
reconstructed $E$ and $B$ maps can be obtained from $Q$ and $U$ using
the reciprocal transformation of the equations~(\ref{eqn:q_eb})
and~(\ref{eqn:u_eb}). The power spectra are then estimated directly
from the Fourier transform of the reconstructed $\widehat{I}$,
$\widehat{E}$ and $\widehat{B}$ maps, by averaging the
$\widehat{a}^X(\mathbf{k})\widehat{a}^{Y*}(\mathbf{k})$ in bins of
width $\Delta k = \Delta l = 20$ (with $X,Y \in \{I,E,B\}$). The
recovered power spectra are then corrected for the smoothing effect
due to the beams, which can be approximated in Fourier space by a
factor $\exp[-l(l+1)\sigma^2]$. However, because of the pixelization
of the maps, this approximation is not good enough. Instead, we have
corrected the power spectra using the power spectrum of the intensity
beam, $B(k) = \left\langle |\widehat{a}^I(\mathbf{k})|^2
\right\rangle$, where $\widehat{a}^I(\mathbf{k})$ is the average of
the intensity beams of the four detectors. This is exact if the beams
are axially symmetric and identical, and otherwise provides a way to
symmetrize the beams in Fourier space. We have used this correction in
all the power spectra shown hereafter.

\begin{figure}
   \centering
   \includegraphics[angle=90,width=8cm]{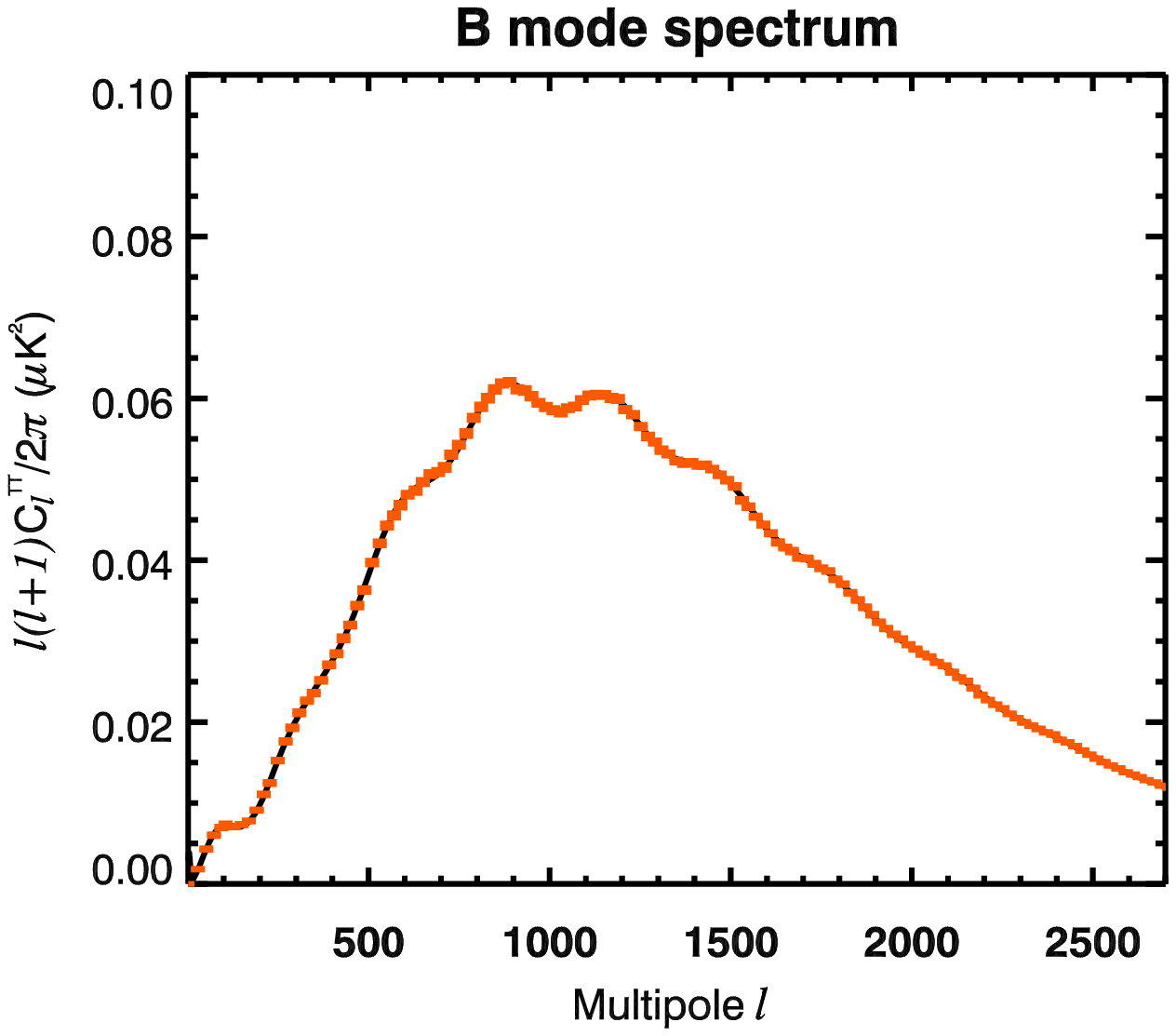} (a)
   \includegraphics[angle=90,width=8cm]{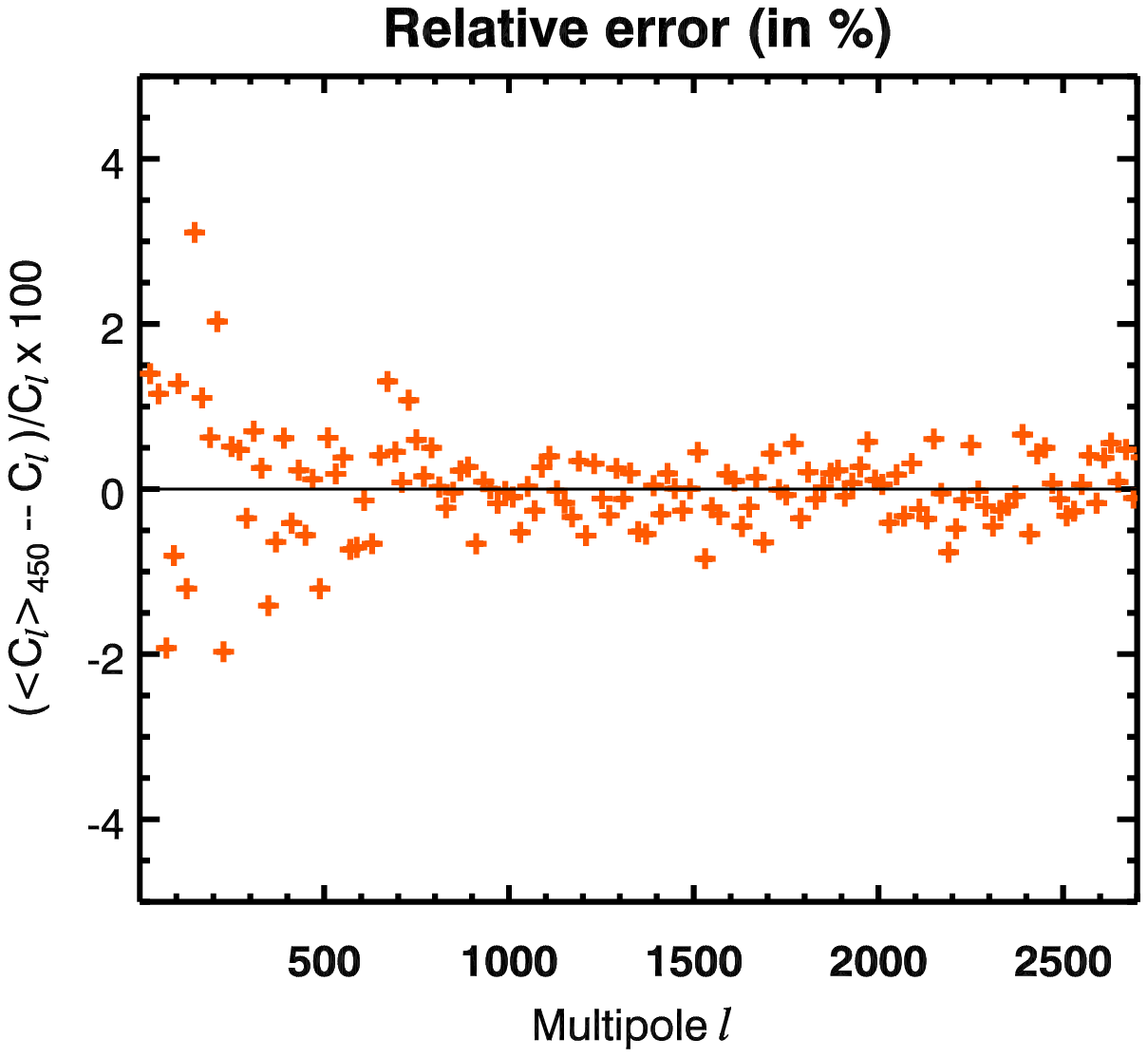} (b)
   \caption{(a) Input and recovered $B$ mode power spectrum with an
   ideal instrument, {\em i.e.} when four identical and axially
   symmetric Gaussian beams are used for both the readout generation
   and the $C_l$ reconstruction. The small peak at $l\sim 100$ is
   produced by the gravitational waves (the tensor to scalar ratio is
   $r=0.1$), while the main pattern peaking at $l\sim 1000$ is due to
   the lensing effect. The error bars are smaller than the thickness
   of the black solid line showing the input model. (b) The relative
   error between the recovered and the initial power spectrum; the
   recovered power spectrum is the average of 450 simultations: the
   statistical error is less than 2\%, thus allowing the detection, in
   non ideal cases, of biases higher than 2\%. Identical figures are
   obtained for $T$, $E$ and $T-E$ correlation power spectra.}
   \label{fig:idealB}
\end{figure}

\begin{figure}
   \centering
   \includegraphics[angle=90,width=8cm]{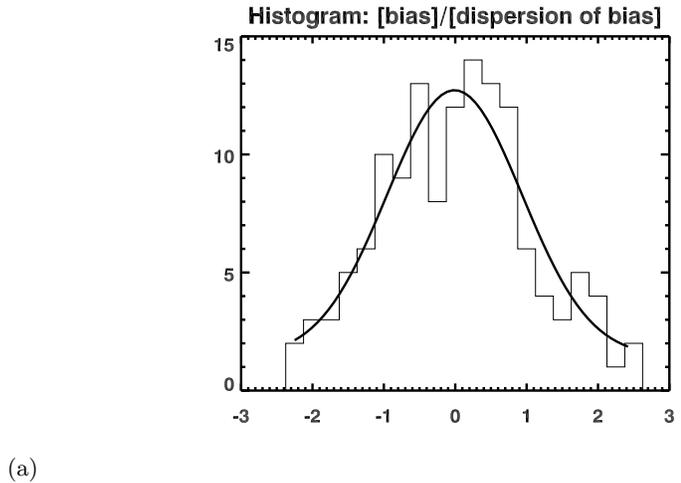}
   \caption{Histogram of the biases divided by the statistical
   dispersion for all multipole bins shown on
   figure~\ref{fig:idealB}. As expected, the histogram is well fitted
   by a Gaussian of unit variance, showing that the dispersion on 450
   simulations gives a good estimate of the errors.}
   \label{fig:idealBhisto}
\end{figure}

The $B$ mode power spectrum reconstructed by using an ideal circular
Gaussian beam in both the readout and reconstruction computations
(assuming Eqs.~(\ref{eqn:quia}) and~(\ref{eqn:quib})) is shown on
Fig.~\ref{fig:idealB}a. The points shown are the average of 450
simulations and the error bars represent the dispersion. The relative
error is shown in Fig.~\ref{fig:idealB}b, demonstrating that the
statistical error on the power spectrum reconstruction averaged over
450 simulations is less than 2\%. Finally, Fig.~\ref{fig:idealBhisto}
presents the histogram of the bias divided by the dispersion, which is
well fitted by a Gaussian with unit dispersion as expected for the
ideal case. Identical results are obtained with other power spectra
($T$, $E$ and $T$-$E$ correlation).

We can now use this tool to estimate the bias induced on the power
spectra reconstruction by the realistic beam shapes described in
section~\ref{sec:problem}.

\subsection{Effects of beams on polarization power spectra}

\begin{figure}
   \centering
   \includegraphics[angle=90,width=8cm]{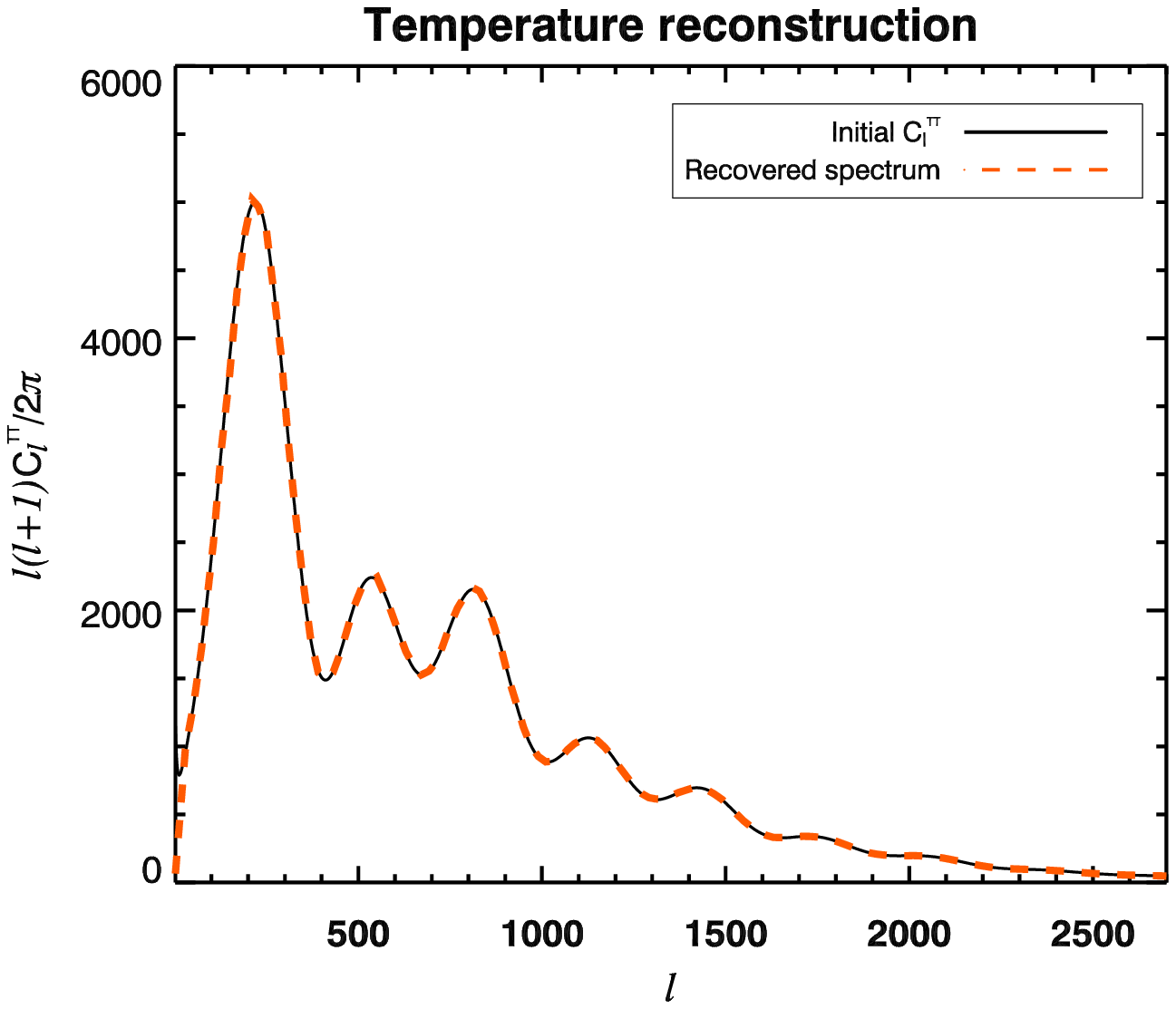}(a)
   \includegraphics[angle=90,width=8cm]{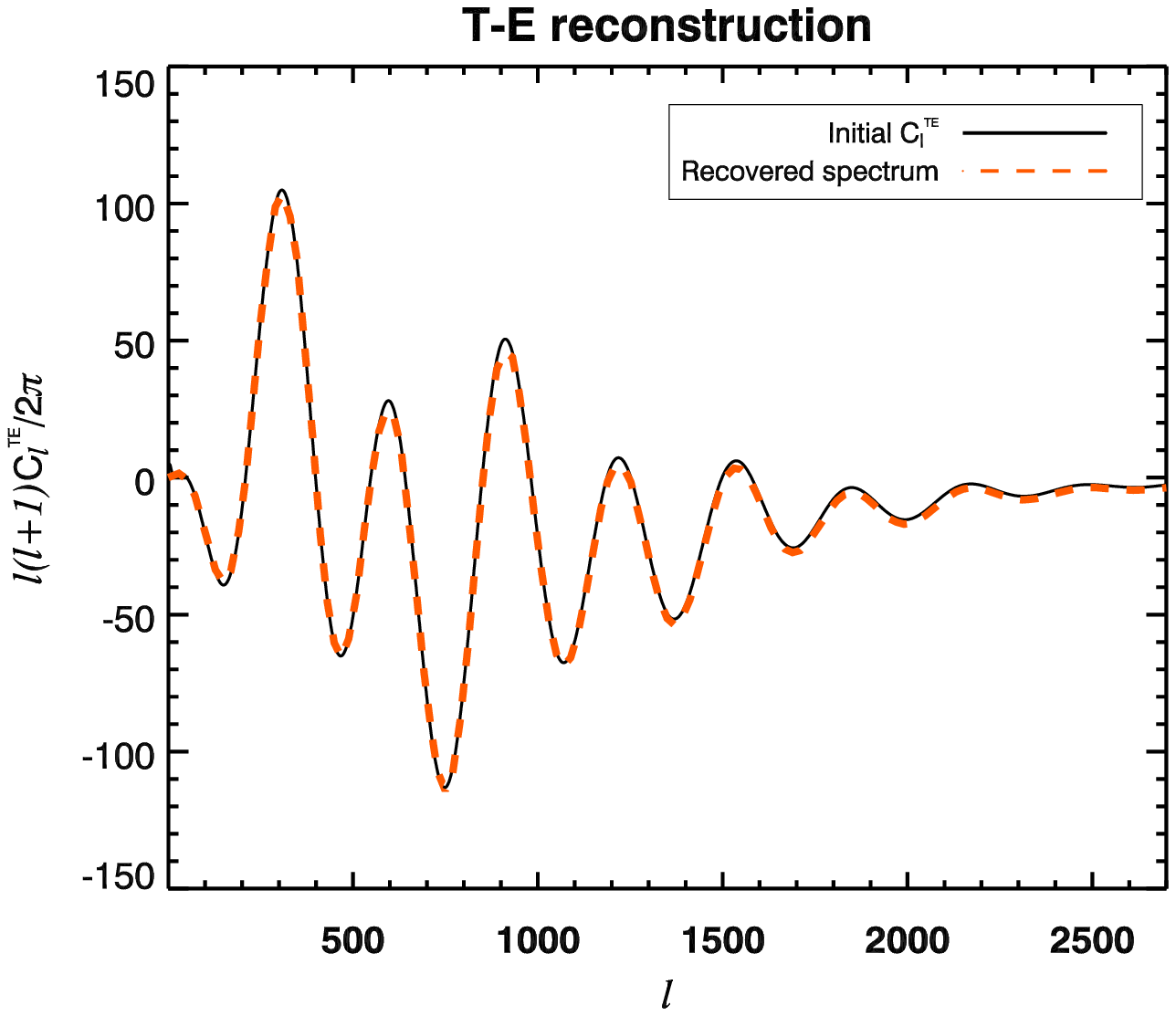}(b)
   \caption{Input and recovered power spectra of (a) temperature and
   (b) $T$-$E$ correlation signals, using the simulated beams of
   Sec.~\ref{sec:problem} for the readout simulation and the $C_l$
   reconstruction. The recovered power spectra are corrected for an
   average symmetric beam effect by multiplying them by the power
   spectrum of the average beam map.}
   \label{fig:nocorr1}
\end{figure}

\begin{figure}
   \centering
   \includegraphics[angle=90,width=8cm]{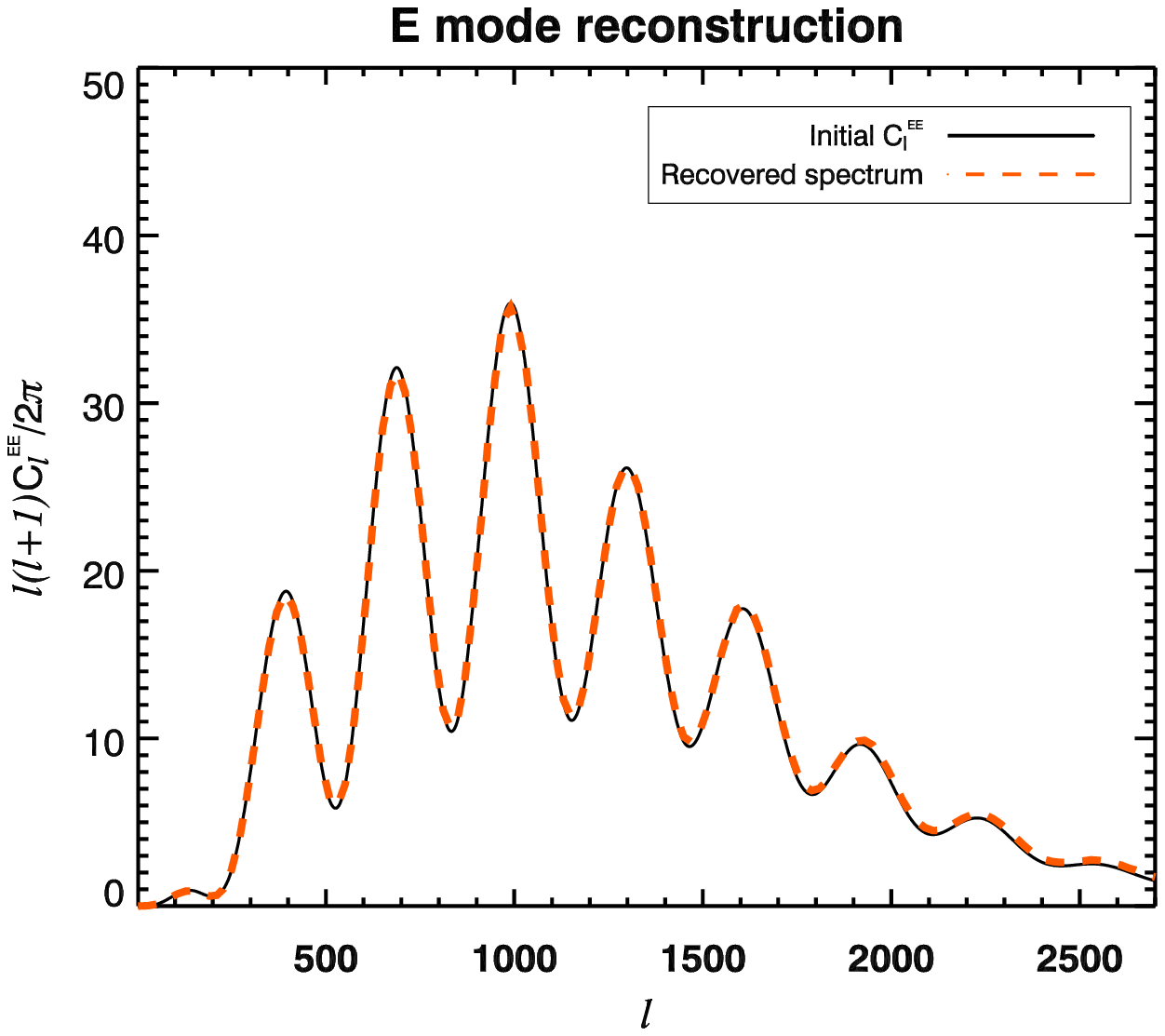}(a)
   \includegraphics[angle=90,width=8cm]{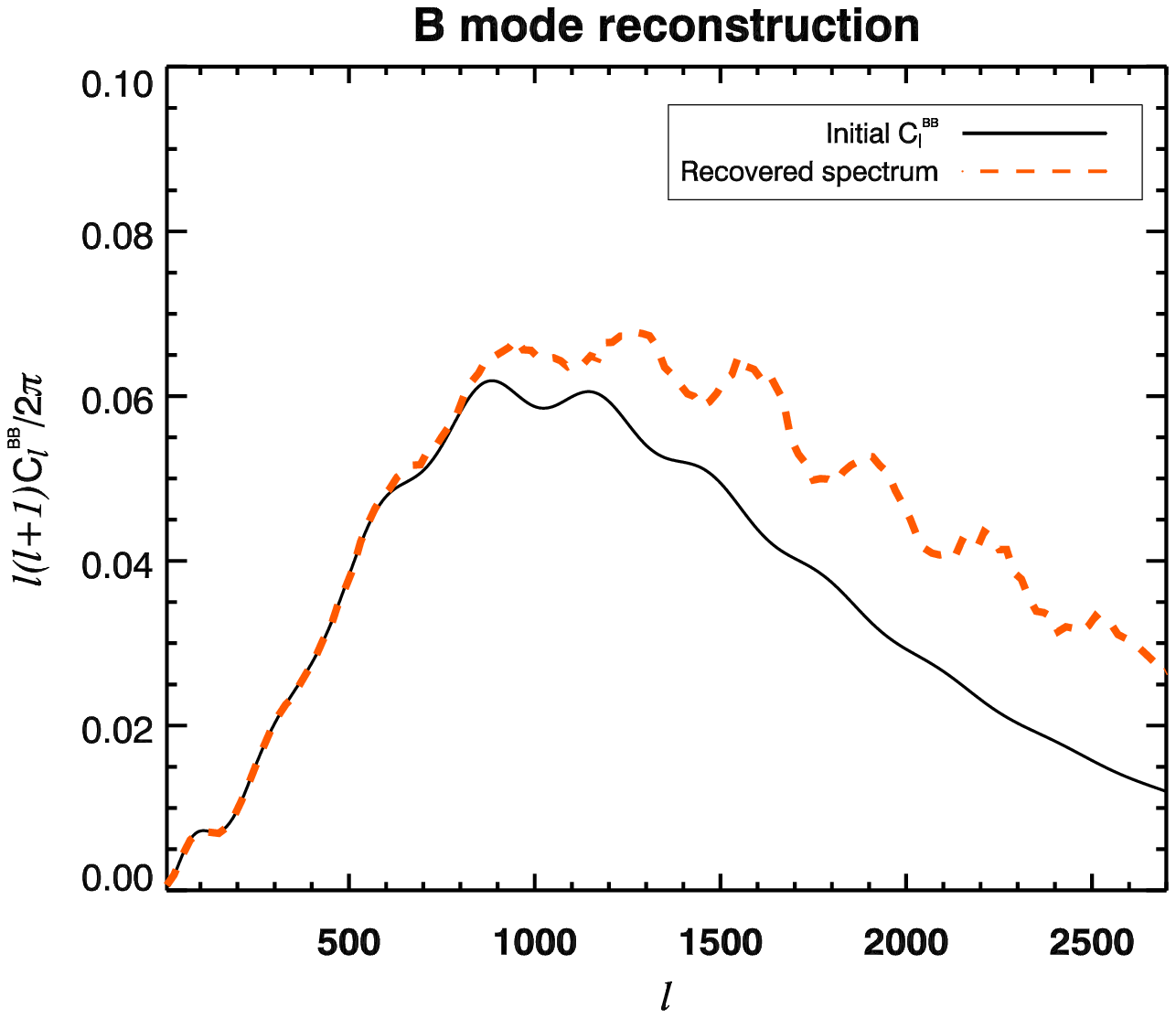}(b)
   \caption{Input and recovered power spectra of (a) $E$ mode and (b)
   $B$ mode signals, using the simulated beams of
   Sec.~\ref{sec:problem} for the readout simulation and the $C_l$
   reconstruction. The recovered power spectra are corrected for an
   average symmetric beam effect by multiplying them by the power
   spectrum of the average beam map.}
   \label{fig:nocorr2}
\end{figure}

We apply our algorithm (both the readout simulation and $C_l$
reconstruction) using the realistic beam patterns $\tilde I$, $\tilde
Q$ and $\tilde U$ presented in section~\ref{sec:problem}. The output
power spectra shown in Fig.~\ref{fig:nocorr1} and~\ref{fig:nocorr2}
are averaged over 450 simulations. The temperature power spectrum is
perfectly recovered, while we can distinguish a small but systematic
excess in the $E$ power spectrum at $l > 2000$ and a systematic loss
in the $T$-$E$ correlation for $l > 1000$. The $B$ mode is strongly
affected after the peak of the lensing signal at $l\sim900$, with a
bias of up to 50\% of the signal at $l\sim 1500$, and about 10\%
around the lensing signal peak at $l\sim 1000$.

\begin{figure}
   \centering
   \includegraphics[angle=90,width=8cm]{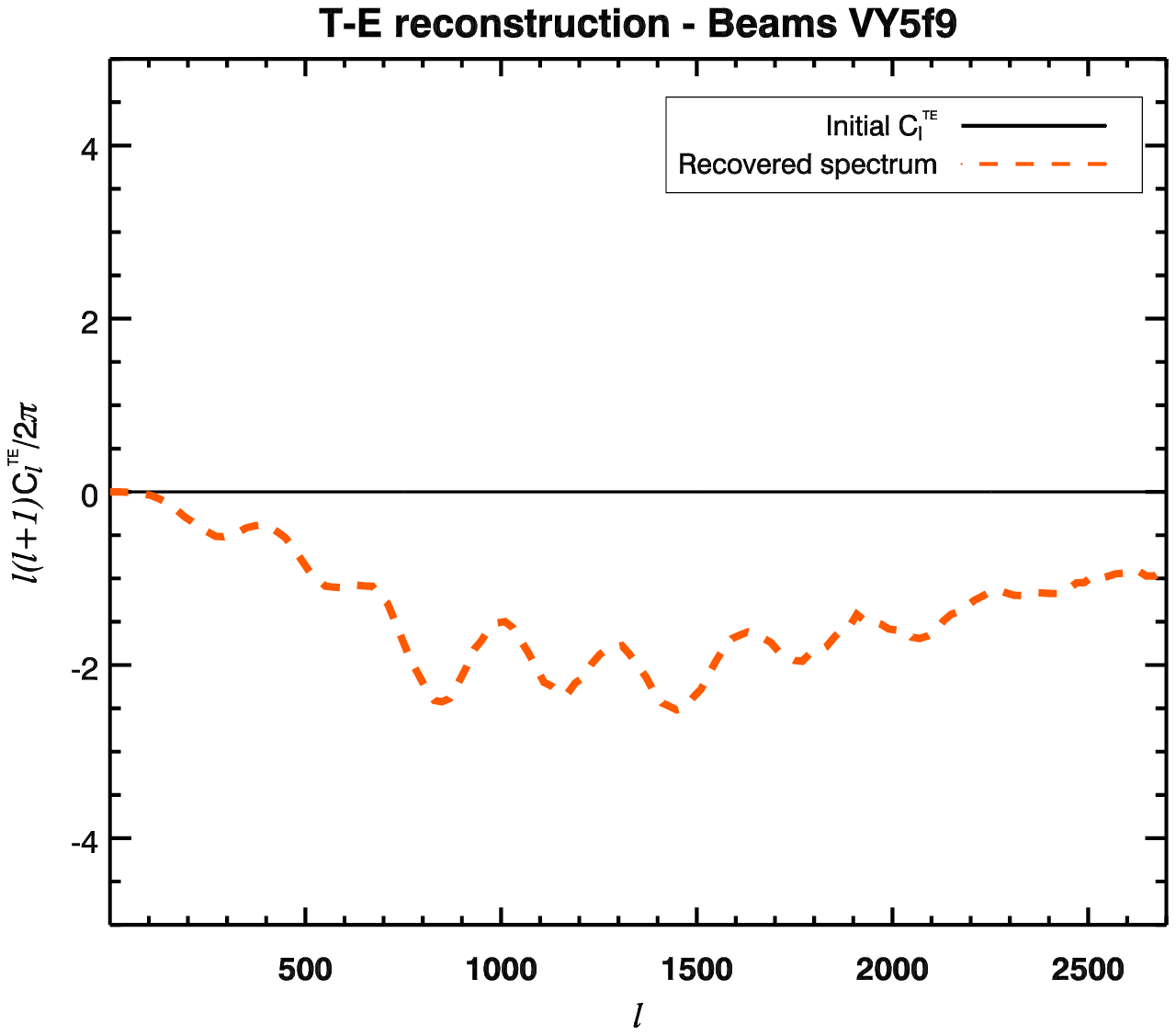}(a)
   \includegraphics[angle=90,width=8cm]{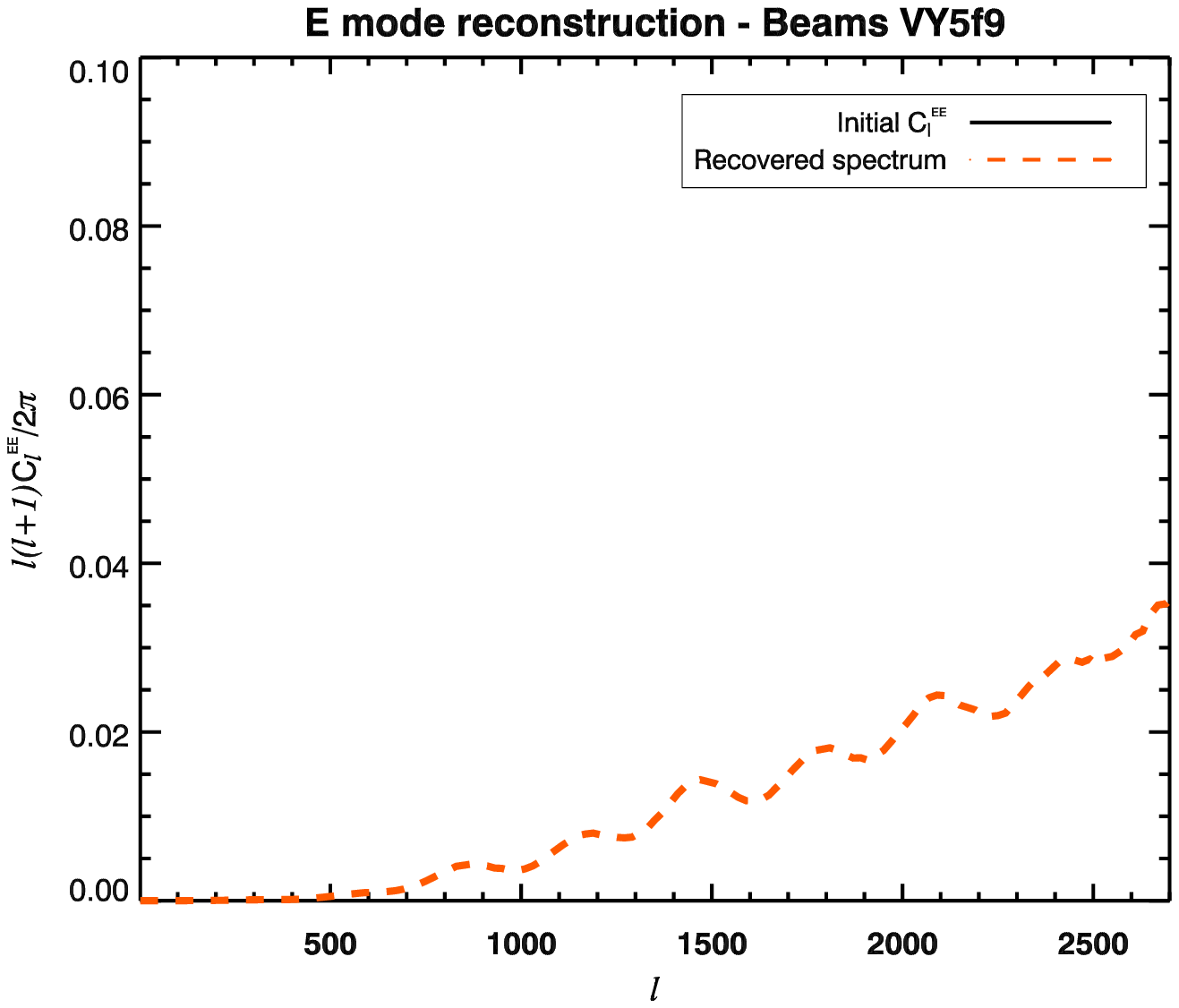}(b)
   \includegraphics[angle=90,width=8cm]{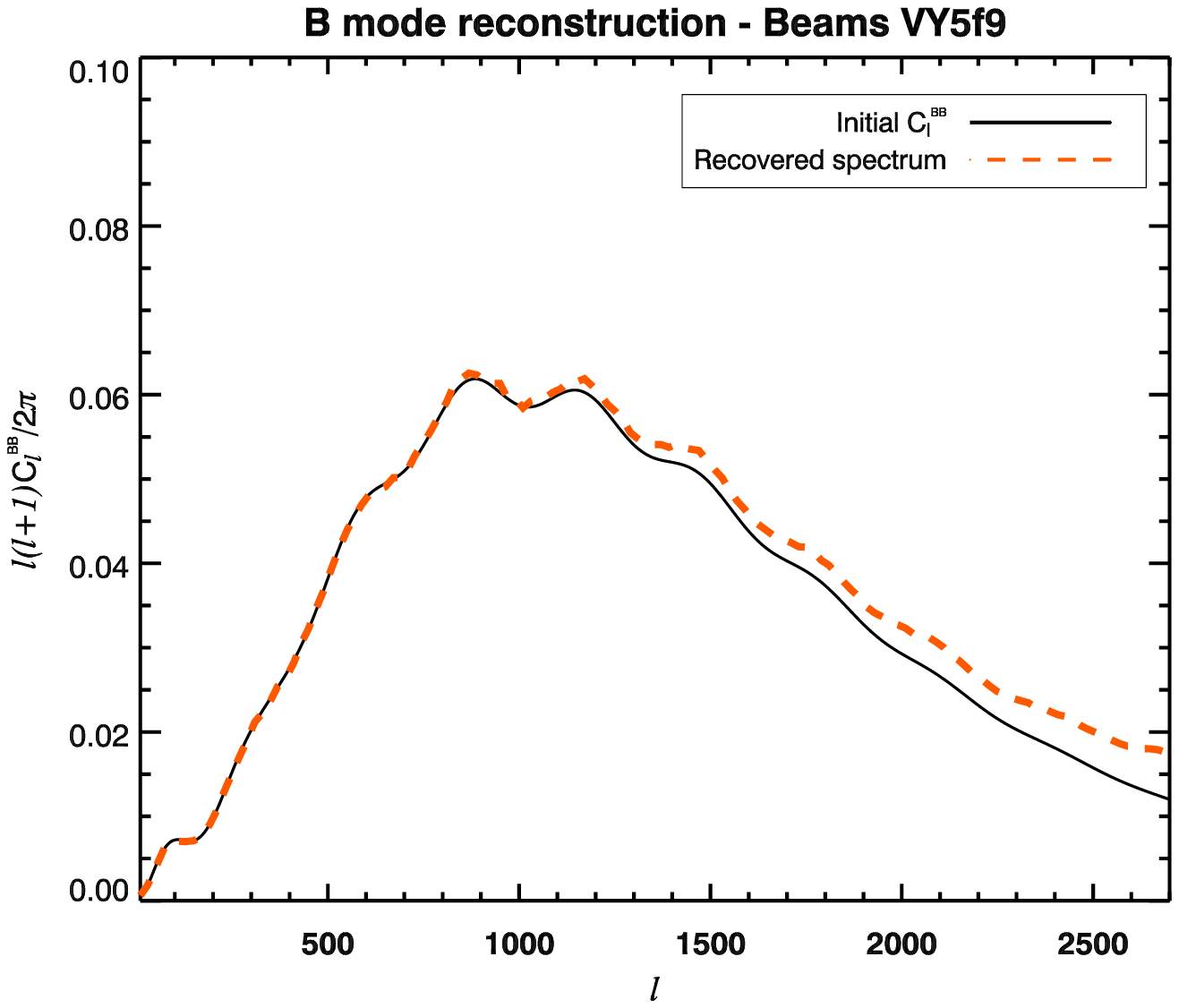}(c)
   \caption{Spurious generation of $E$ and $B$ modes from temperature
   signals using simulated beams of Sec.~\ref{sec:problem} for the
   readout simulation and the $C_l$ reconstruction, but with no
   initial $E$ mode.}
   \label{fig:nocorrnoe}
\end{figure}

The spurious $B$ mode may come from leakage of either the temperature
or the $E$ mode. In order to separate the two possible origins, we
have done the same simulation using the realistic beams $\tilde I$,
$\tilde Q$ and $\tilde U$ when computing both the readouts and $C_l$
reconstruction but with no input $E$ mode (and no $T-E$ correlation).
The results for $T$-$E$, $E$ and $B$ power spectra are shown on
Fig.~\ref{fig:nocorrnoe} (the temperature power spectrum is not
modified). We observe that the spurious $B$ mode is about three times
smaller in this case, indicating that about 2/3 of the spurious $B$
seen with realistic input $E$ mode came from $E$ leakage. On the other
hand, the level of the $E$ mode is much higher than would be expected
if it came from a mixing from $B$ modes into $E$ modes.

In order to check this, we made a simulation with elliptic
Gaussian beams identical for detectors within the same horn, but with
different ellipse directions for different horns. We have assumed
Eqs~(\ref{eqn:quia}) and~\ref{eqn:quib} to represent the ideal
polarization sensitivity of the channels, so that there is no total
intensity leakage into polarization signal due to beam difference,
and, again, used no input $E$ mode. In this situation, the recovered
$E$ mode is a small fraction of the input $B$ mode signal, {\em i.e. }
much smaller than the recovered $E$ mode in the previous test. This
means that the $E$ mode recovered in the previous test (using
realistic beams) was not due to a leakage from $B$ modes to $E$ modes,
but rather from a leakage from $T$ to $E$. We conclude that, both the
$E$ mode and the spurious $B$ mode found with realistic simulated
beams when there is no $E$ mode in the readout simulation, come from a
temperature leakage due to the differences in the beam patterns
between detectors within the same horn, which is up to 0.9\%
(Fig.~\ref{beam_difference}a).



\begin{figure}
   \centering
   \includegraphics[width=8cm]{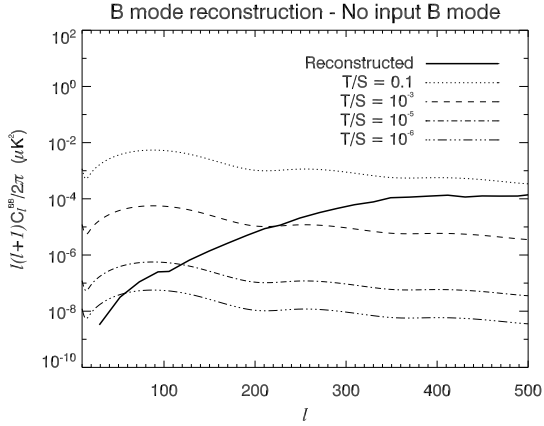}
   \caption{Recovered $B$ mode power spectrum in a simulation with no
initial $B$ mode. The theoretical $B$ mode power spectra due to
primordial gravitational waves are also shown for different values of
the tensor-to-scalar ratio : 0.1, $10^{-3}$, $10^{-5}$ and $10^{-6}$
from top to bottom.}
   \label{fig:pslowl}
\end{figure}

We thus see that two different effects produce the observed spurious
$B$ mode. First, there is a mixing between the two polarization modes,
essentially from $E$ to $B$ as $E \gg B$ on all scales, due to the
beam mismatch between the two different horns. Second, there is a
temperature leakage, this time due to the beam mismatch between the
PSB within the same horn.

As seen on Fig.~\ref{fig:nocorr2}, the beam asymmetry affects mainly the
high-$l$ part of the power spectra (typically $l > 500$). However,
it is not negligible at low $l$, where the gravitational waves lie.
The Fig.~\ref{fig:pslowl} shows the recovered power spectra when there
is no initial $B$ mode in the simulation compared to the expected $B$ mode
signal from gravitational for various tensor-to-scalar ratios. The leakage
from $T$ and $E$ mode to $B$ mode power spectrum becomes greater than
the gravitational wave $B$ mode signal for $T/S \lesssim 10^{-5}$.

\subsection{Link with previous work}

Various studies have been done on the systematic effects on CMB
polarization measurements, in particular the exhaustive work by Hu~{\em
et al.}, \cite{Hu2002} (referred to as HHZ hereafter). This paper
tries to estimate analytically the systematic effects on $B$ mode
power spectrum, using a second order expansion and relating the
terms of the expansion to beam defects such as, for example,
pointing error, ellipticity, monopole leakage or calibration. The
different systematic effects are assumed to be described by a
statistically isotropic field, with a power spectrum of the form :
\begin{equation}
\label{eqn:pssyst}
C_l \propto \exp (-l(l+1)\alpha^2),
\end{equation}
so that the leakage from $T$ or $E$ to $B$ can be written as a
convolution between $EE$ or $TT$ and systematic effect power spectra
(Eqs.~30 and~34 in HHZ).

In our approach, the $I$, $Q$ and $U$ maps are given through the
signals of four detectors (Eqs.~\ref{eqn:reciqu}). In the quasi-ideal
case of all $Q$ and $U$ beams defined by Eqs~\ref{eqn:quia}
and~\ref{eqn:quib}, but with intensity beams different between the two
horns, it can be shown that the error on the $B$ mode power spectrum
is given by :
\begin{equation}
\label{eqn:dbb}
\delta C_l^{BB} = \frac{\left\langle \left| \Delta
\widetilde{I}_{13}(\mathbf{l})\right|^2 
\cos^2 (2\phi_l) \sin^2(2\phi_l) \right\rangle_{\phi_l}}{B_l}
C_l^{EE}(\sigma)
\end{equation}
where $\Delta \widetilde{I}_{13} = \widetilde{I}_1 - \widetilde{I}_3$
is the beam difference, $B_l$ is the average beam power spectrum and
$C_l^{EE}(\sigma)$ is the power spectrum of the $E$ map convolved with
the average beam. This form is a particular case of the one given by
HHZ, and describes completely the systematic effects on the $B$ mode
power spectrum due to the beam mismatch between horns. Note however
that there is no mixing between different $l$ of the power
spectra. The major difference with HHZ estimate is in the shape of the
beam difference power spectrum (first factor in Eq.~\ref{eqn:dbb})
which can be fitted by $C_l^{\delta I} \propto l^4$ in the interval $l
\in \{1,\ldots,3000\}$, very different from HHZ assumption,
Eq.~\ref{eqn:pssyst}.

In the particular case we consider (beam mismatch) our approach gives
more realistic results, as we use accurate simulations of the
beams. Moreover, the power spectrum of the defect due to beam mismatch
we compute from these beams is very different to HHZ assumption,
leading to a different estimate of the size of the effect.

\section{Correction of polarization spectra for systematic errors}
\label{sec:correction}
We shall now propose a simple way to correct for the spurious $B$ mode
deduced from the observations of the previous section. The idea is to
assume that temperature and $E$ mode maps are recovered well enough to
estimate the $T$ and $E$ to $B$ mode leakage, if we know the beam
patterns. We will discuss three cases of $C_l$ correction, depending
on the knowledge of the beams. In all three cases, the initial
readouts are generated with the realistic $\tilde I$, $\tilde Q$ and
$\tilde U$ beam patterns of Sec.~\ref{sec:problem}.

\subsection{Perfect knowledge of the intensity beam pattern}

In order to have an idea of the ability of the method to remove the
spurious $B$ mode, we have tested it in the case of a perfect
knowledge of the intensity beam patterns. However, because of the lack
of polarized point sources with known polarization characteristic, we
assumed that only the intensity beam patterns $\widetilde I$ were
perfectly measured while the $\tilde Q$ and $\tilde U$ needed for the
$C_l$ correction are computed using relations~(\ref{eqn:quia})
and~(\ref{eqn:quib}) with the relevant $\tilde I$ in all the three
cases considered.

The method is as follows. By a quick analysis of the data, we
would find maps and their corresponding power spectra similar to the
ones shown in Figs.~\ref{fig:nocorr1} and~\ref{fig:nocorr2}. Since
the $T$ and $E$ mode power spectra are recovered with a very good
approximation, we may assume that the recovered maps are good as
well. Starting with the temperature and the $E$ mode maps, assuming
no initial $B$ mode and using a precise knowledge of the beams, we
could then simulate the instrument signals. From the previous
consideration, we expect to find, from these simulated signals, a
spurious $B$ mode polarization coming both from a temperature leakage
and a polarization mode mixing. The $B$ mode power spectrum of this
simulated signal should be an estimate of the spurious $B$ mode.

\begin{figure}
  \centering
  \includegraphics[angle=90,width=8cm]{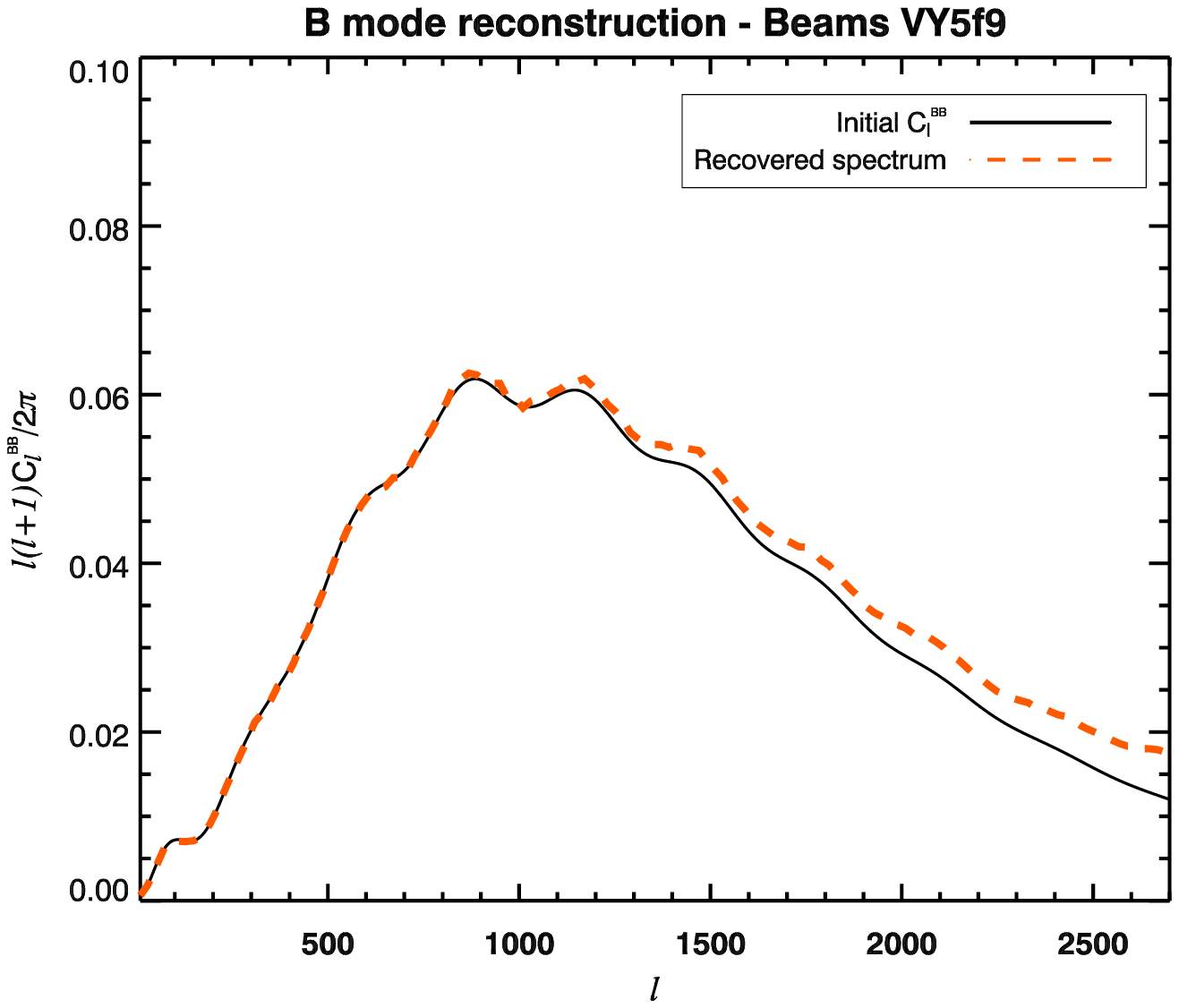}
  \includegraphics[angle=90,width=8cm]{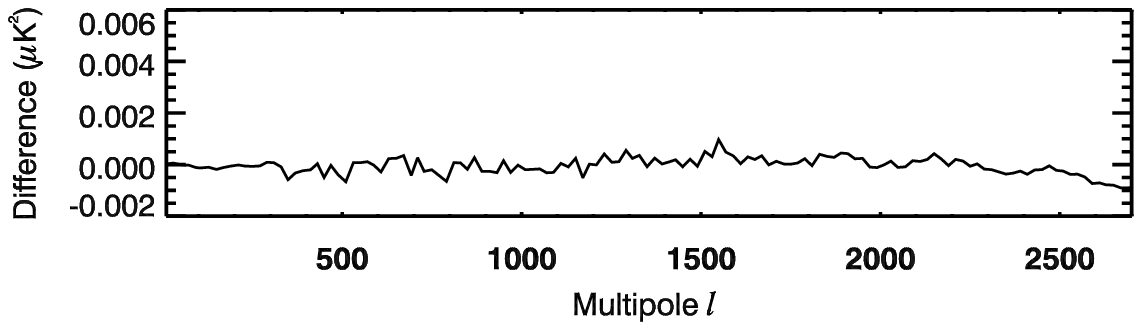}
  \caption{Recovered $B$ mode power spectrum before (red, dashed line)
  and after (green, dot-and-dashed line) correction. The power
  spectrum is corrected by subtracting the estimated leakage (blue
  dotted line) assuming knowledge of the exact beams (see text). {\em
  Bottom:} Difference between corrected and initial power spectra.}
  \label{fig:correxact}
\end{figure}

The result for the $B$ mode correction, using exact $\tilde I_a$ and
$\tilde I_b$ and assuming relations~(\ref{eqn:quia})
and~(\ref{eqn:quib}) for the leakage estimation, is shown on
Fig.~\ref{fig:correxact}. The correction allows us to reduce the bias
down to less than 1\% of the lensing signal in the interval $2 < l <
1500$.

\subsection{Assuming identical beams within the same horn}

In a second case, we supposed that, in order to increase the signal to
noise ratio, we need to use the signal from both detectors within one
horn to measure the beam patterns. With this method, we would find as
beam pattern the average of the beams of the two detectors within one
horn, {\em i.e.} the average error on the beams is about 0.5\% of the
beam maximum. 

\begin{figure}
  \centering
  \includegraphics[angle=90,width=8cm]{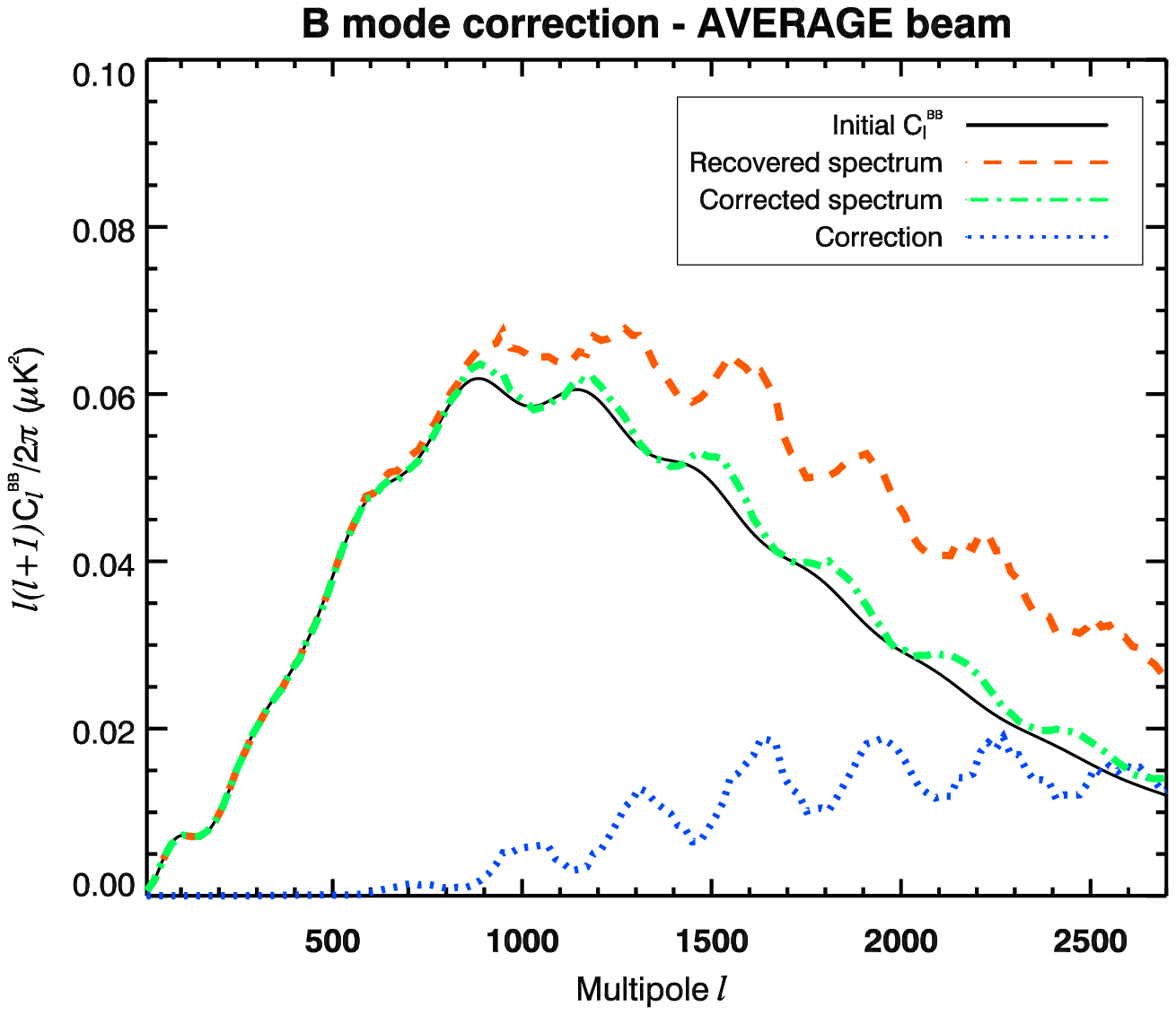}
  \includegraphics[angle=90,width=8cm]{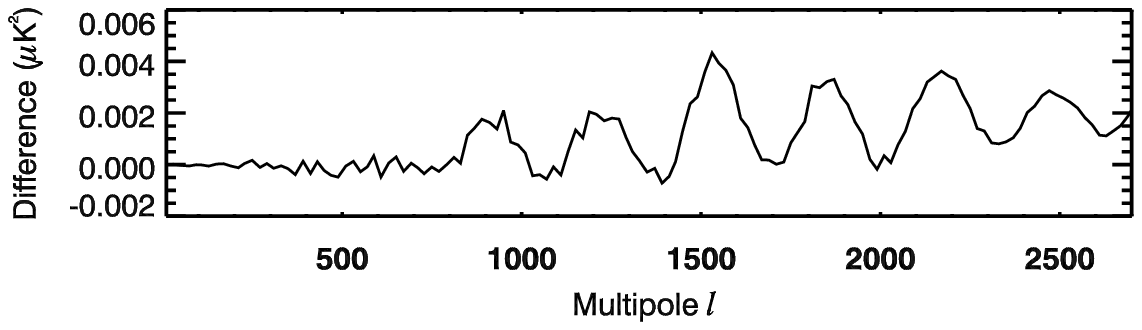}
  \caption{Recovered $B$ mode power spectrum before (red, dashed line)
  and after (green, dot-and-dashed line) correction. The power
  spectrum is corrected by subtracting the estimated leakage (blue
  dotted line) using beams averaged within one horn (0.5\% error, see
  text). {\em Bottom:} Difference between corrected and
  initial power spectra.}
  \label{fig:corrmean}
\end{figure}

The result obtained for the $B$ mode correction is presented on
Fig.~\ref{fig:corrmean}. This time, there is still some bias left in
the corrected power spectrum, around 3\% at $l\sim 1000$ and up to
13\% for $l \sim 2100$.

\subsection{Fitting the beams with elliptic Gaussians}

If we have only few point sources or low signal-to-noise ratio on
signal, we may want to parametrize the beam patterns with a function
requiring a small number of parameters. As an example, we have fitted
the four intensity beam patterns by elliptic Gaussian. The error of
the fit is around 2\% of the maximum of the beam. 

\begin{figure}
  \centering
  \includegraphics[angle=90,width=8cm]{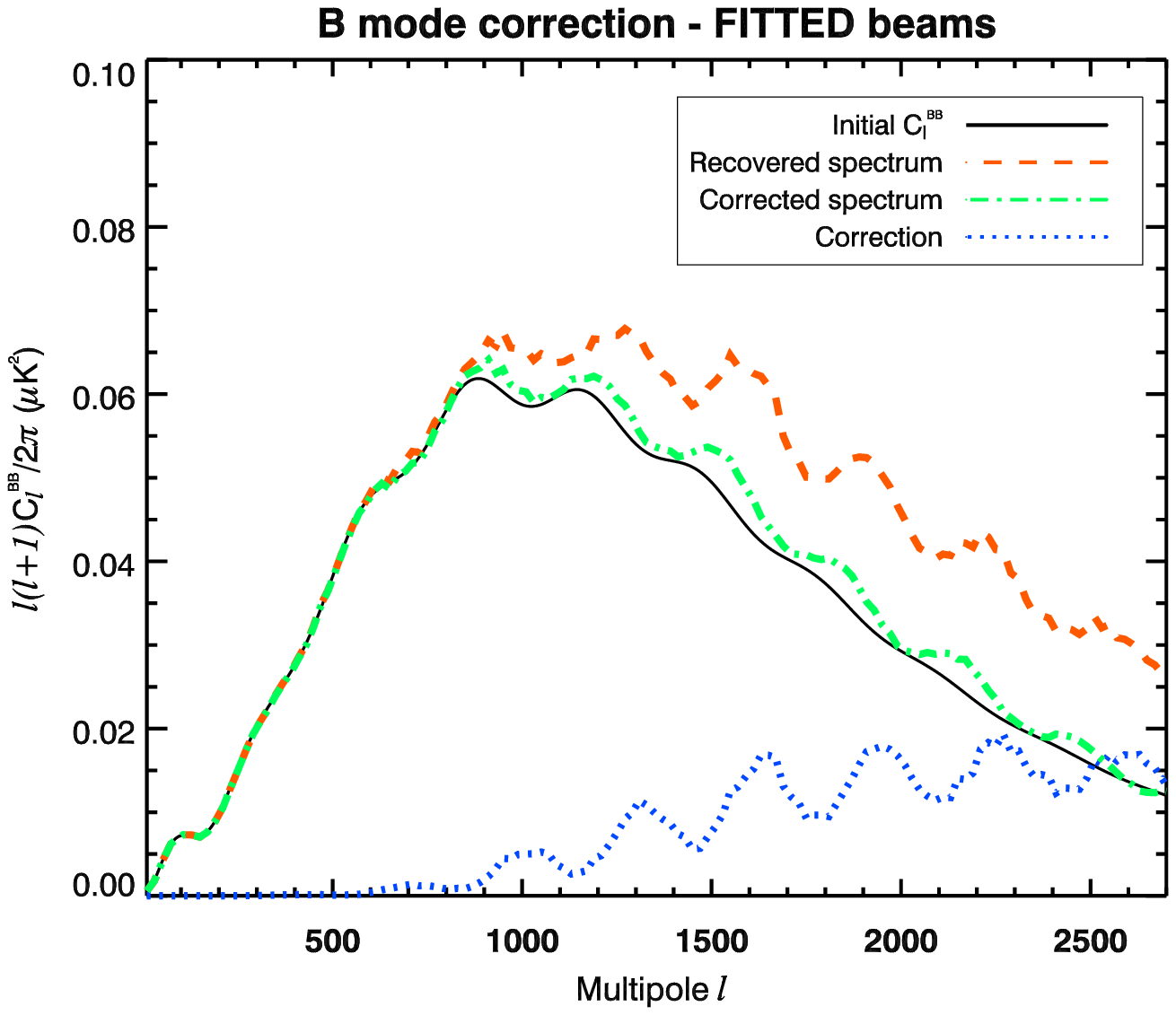}
  \includegraphics[angle=90,width=8cm]{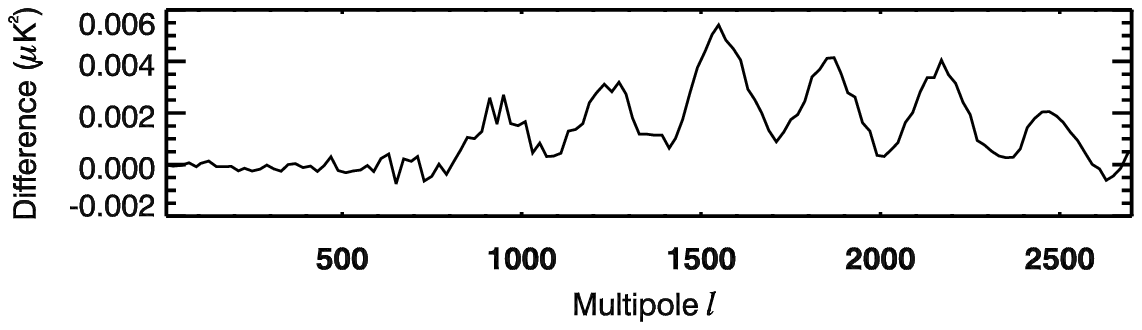}
  \caption{Recovered $B$ mode power spectrum before (red, dashed line)
  and after (green, dot-and-dashed line) correction. The power
  spectrum is corrected by subtracting the estimated leakage (blue
  dotted line) using elliptic Gaussian beams fitted on the exact beams
  (2\% error, see text). {\em Bottom:} Difference between corrected and
  initial power spectra.}
  \label{fig:corrfit}
\end{figure}

The result is shown in Fig.~\ref{fig:corrfit}, together with the
difference between the corrected and initial $B$ mode power
spectra. The result is very similar to that of Fig.~\ref{fig:corrmean}
(using horn-averaged beams), though the remaining bias is slightly
higher.

This simple method thus seems efficient to recover the right height
of the lensing effect peak at $l\sim 1000$. Though it is applied here
in the case of a simple scanning strategy (parallel scans), it should be
applicable to any scan strategy, as soon as the bias estimation is
done using the beams as precise as possible and the same scanning
strategy as the real one.

\section{Conclusions}
\label{sec:conclusions}

In this paper, we have shown the effect of asymmetric telescope beams
on the bolometric measurements of polarization of incoming radiation
by considering the case of the Planck satellite mission. We have used
electromagnetic simulation of the optical system (including telescope
and horns) to compute the main beam shapes of the different detectors
of Planck. These beams are roughly Gaussian elliptical, with a major
axis 10\% larger than the minor axis and with essentially different
orientations of the beam ellipses for the two horns to be combined to
measure the full set of Stokes parameters, $I$, $Q$ and $U$.

By simulating the scan of a patch of the sky by Planck with these
realistic, simulated beams, we have estimated the bias induced on the
$E$ and $B$ mode polarization spectra due to their asymmetric
shapes. We first remark that the $E$ mode power spectrum is very well
recovered (once corrected for an effective symmetric beam), the bias
being around 0.1\% of the signal in the multipole range $300<l<2000$,
where lies the most interesting part of the signal. On the other hand,
the $B$ mode is affected by a bias around 10\% at the peak of the
lensing signal ($l\sim 1000$) and increasing for higher $l$, up to
100\% of the signal at $l\sim 2500$. This bias has two origins. First,
it is produced by the difference of beam patterns of two different
horns combined to measure $Q$ and $U$. This difference induces mainly
an error on the polarization angle, which turns to a mixing of $E$ and
$B$ modes. Since, in general, $E \gg B$, we observe finally a leakage
from $E$ to $B$. The second origin of the bias is the minor difference
of beam patterns of two PSB channels with orthogonal polarizations
within the same horn, which induces a temperature to polarization
leakage.

Finally, we have proposed a way to correct the $B$ mode power spectrum
from the above bias in a one-step correction which uses the measured
$T$ and $E$ maps to compute the expected leakage into $B$ when they
are observed with a model of the instrument's beams. The efficiency of
this correction depends on the precision of the beam knowledge: for
example, using elliptical Gaussian fits of the actual beams allows us
to reduce the bias from 10\% to 3\% at the lensing signal peak, $l\sim
1000$. In all cases, this first order correction has been shown to
reduce significantly $B$ mode contamination. More refined treatments,
currently being investigated, are expected to be yet more efficient if
needed.

\begin{acknowledgements}
      The authors would like to thank Ken Ganga for fruitful
      discussions. We also want to thank the referee for interesting
      remarks which helped us to improve the manuscript. This work was
      supported by the Ulysses 2002 and 2003 Research Visit Grants
      paid by the Enterprise Ireland and CNRS France.
\end{acknowledgements}


\appendix

\section{Simulation of the Planck HFI telescope beams} 
\label{sec:beams}

It will not be possible to measure HFI beams on ground. The HFI
bolometers indeed work only at 100 mK and are designed for the
thermal load of a few Kelvin environment in space. Whereas the
instrument can be put in a large vacuum tank cooled to 4 K, it is not
possible to perform far field measurements of the full system, which
would require placing a source hundreds of meters away from the
instrument. Hence, responses can only be measured at subsystem level
(e.g. bolometers + horns) and must be associated to a physical model
of the telescope to predict the beam shape of the complete integrated
optical system.

In this paper, for the investigation of beam mismatch effects, we use
computer-simulated Planck HFI beams. We consider four HFI-143 beams
comprising eight PSB channels used for the polarization measurements
in the band centered at the frequency $\nu=143$ GHz. The polarization
direction of each PSB is specified by the polarization angle $\psi$ on
the sky and labeled by the relevant index of the channel ($\psi_{1a}$,
$\psi_{1b}$ etc) as shown in Fig.~\ref{fig:planck-scanning-QU}(b).

In the design of the focal plane unit (FPU), the polarization angles
$\psi_{i\alpha}$ notifying the channels are measured from the upward
direction of local meridian of the spherical frame of spacecraft (SC)
having the geometrical spin axis of telescope as a pole and counting
the angles $\Psi$, as viewed from telescope,
clockwise toward the direction of maximum polarization sensitivity of
the channel. Similarly, we define the polarization angle $\psi$ at
each observation point $\vec x$ in the beam pattern.  The direction of
maximum polarization sensitivity is the major axis of polarization
ellipse at point $\vec x$; for the angles $\psi_{i\alpha}$ notifying
the channels, $\vec x$ is the beam axis defined as the point of
maximum intensity $\widetilde I$ (at this point, the beam field
is linearly polarized).

Following this definition, we consider eight PSB channels of the HFI-143 
beams which are sensitive to the linearly polarized radiation 
with polarization angles on the sky $\psi_{1a}=\psi_{2a}=45^{\circ}$, 
$\psi_{1b}=\psi_{2b}=135^{\circ}$, $\psi_{3a}=\psi_{4a}=0^{\circ}$, and 
$\psi_{3b}=\psi_{4b}=90^{\circ}$. 
The four beams are arranged in two pairs (1 and 3, 2 and 4), with 
two beams of each pair scanning the sky along the same scan path 
as shown in Figs.~\ref{fig:planck-scanning-QU} and~\ref{fig:focplane}.
In each pair of beams, the angles $\psi_{i\alpha}$ correspond to a 
full set of four PSB detectors for polarization measurements with
optimized polarimeter configuration (Couchot et al, \cite{Couchot}).

The power patterns of two beams tracing the same scan path,
HFI-143-2-a/b and HFI-143-4-a/b, are shown in Fig.~\ref{beams} as
projected on the sky in the spherical frame SC with coordinates
$\varphi_{SC}, \theta_{SC}$ ($\eta_{SC}=90^{\circ}-\theta_{SC}$) where
the azimuthal angle $\varphi_{SC}$ is counted to the right from the
optical axis of telescope as viewed from the sky and the polar angle
$\theta_{SC}$ is measured from the upward direction of nominal spin
axis (the optical axis of the telescope corresponds to
$\varphi_{SC}=0^{\circ}$ and $\eta_{SC}=5^{\circ}$).

Notice that both the $a$,$b$ labels of channels and polarization angles 
$\psi$ are conventionally defined with respect to meridians (verticals) of 
the SC frame viewed from telescope, as accepted by the FPU design. 
In the meantime, because of the scanning strategy, the reference axes 
for the Stokes parameters on the CMB maps are usually parallels 
(horizontals) of the SC frame viewed from the sky to the telescope. 

To reconcile these definitions, we continue to use the polarization 
angles $\psi$ and the established notations $a$,$b$ for the PSB channels. 
In the same time, for processing the readouts according to Eqs. 
(\ref{eqn:sa})--(\ref{eqn:sigconva}), we define beam responses 
${\widetilde I}$, ${\widetilde Q}$, ${\widetilde U}$, ${\widetilde V}$ 
in SC frame viewed from sky, with the first and the second reference 
axes being the azimuth $\varphi_{SC}$ and the elevation $\eta_{SC}$, 
respectively (they constitute the right-hand frame $xy$ for defining 
Stokes parameters on the sky as viewed from sky to telescope). 
With these definitions, the polarization angle in $xy$ frame is the 
angle $\alpha$ in Eqs. (\ref{eqn:sa}), (\ref{eqn:sb}) where 
$\alpha=\psi+90^{\circ}$.

The beams in Fig.~\ref{beams} are computed with an extended version of
the fast physical optics code (Yurchenko et al, \cite{FPO}) developed
specifically for the efficient simulations of the Planck HFI
beams. The extended code allows us to propagate via the telescope the
aperture field of the HFI horns mode--by--mode at various
frequencies. The aperture field is generated by the PSB bolometers
considered as polarized black--body radiators (in the transmitting
mode) located at the rear side of the horns. In this way, we obtain
the band--averaged far-field patterns of Stokes parameter responses
$\widetilde I$, $\widetilde Q$, $\widetilde U$, $\widetilde V$ of the
broad--band telescope beams as produced by the actual corrugated
horns (Yurchenko et al, \cite{ESTEC}) rather than by
simplified model feeds.

Rigorous computations of beams require scattering matrix simulations of horns 
(Murphy et al, \cite{horns}). In this approach, the effective modes of the 
electric field at the horn aperture, $E_{nm}$, are represented via the 
canonical TE, TM modes ${\cal E}_{nj}$ of a cylindrical waveguide as follows
\begin{equation}
E_{nm}(\rho,\varphi) = \sum_{j=1}^{2M} S_{nmj}{\cal E}_{nj}(\rho,\varphi)
\end{equation}
where $S_{nmj}$ is the scattering matrix computed by Murphy et al
(\cite{horns}) for each horn at various frequencies, 
$n= 0, 1, ..., N$ is the azimuthal index and $m,j= 1, 2, ..., 2M$ 
are the radial indices accounting for both the TE ($m,j= 1, ... , M $) 
and TM ($m,j= M+1, ... , 2M$) modes.

Recent simulations of the HFI-143 beams (Yurchenko et al, \cite{Fit})
were performed with the scattering matrices of size $20 \times 20$
($M=10$, $N=1$) using nine sampling frequencies spanning the band
$\nu=123-163\,$GHz. Although the power patterns of these beams only
slightly differ from those computed earlier (Yurchenko et al,
\cite{ESTEC}) with matrices $10 \times 10$ and five sampling
frequencies ($\Delta P < 0.1\;$dB at $P=-3\;$dB and $\Delta P <
1.5\;$dB at $P=-30\;$dB), the effect on the difference between the
beams of different polarization and on the fine polarization
properties of beams is noticeable.  This suggests that the latter
parameters could be sensitive to other features of the model as well.

In this paper, the HFI-143 beams are computed with two essential
updates compared to (Yurchenko et al, \cite{Fit}): the horn design is
slightly altered so that the horns are now slightly elongated compared
to those used earlier, and the horn positions are now the final ones,
being defined by the parameter $R_C=1.2$ mm that specifies the refocus
of the horn aperture with respect to the geometrical focus of
telescope for each beam.

The horn positions were optimized to achieve the best resolution 
(the minimum beam width) of the broadband beams (this also maximizes 
the gain) so that the value $R_C=1.2$ mm is close to the optimal horn 
positioning. We use updated scattering matrices of size $20 \times 20$ 
for representing the horn field and nine sampling frequencies for 
spanning the frequency band $\nu=121-165$ GHz which is characteristic 
of the updated horns.

At this stage, we assume smooth telescope mirrors with ideal
elliptical shape, perfect electrical conductivity of their reflective
surfaces, and with ideal positioning of both the mirrors and horn
antennas. The convergence accuracy of computations was better than
$0.1\%$ relative to the maximum of the beam intensity pattern
$\widetilde I(\varphi,\eta)$ (for comparison, the difference of
the broadband power pattern and the pattern computed at the central
frequency is about $1\%$).

To be perfectly representative of what the actual beams may be, one 
should take into account possible misalignments of the optics, tilts 
and deformations of the mirrors, etc. In principle, however, tolerances 
on the alignment of mirrors and positioning of horns are such that the 
modifications they induce on the beams are supposed to be small, and we 
neglect this last issue for the present work.

   \begin{figure}[tb]
   \centering
   \includegraphics[width=8cm]{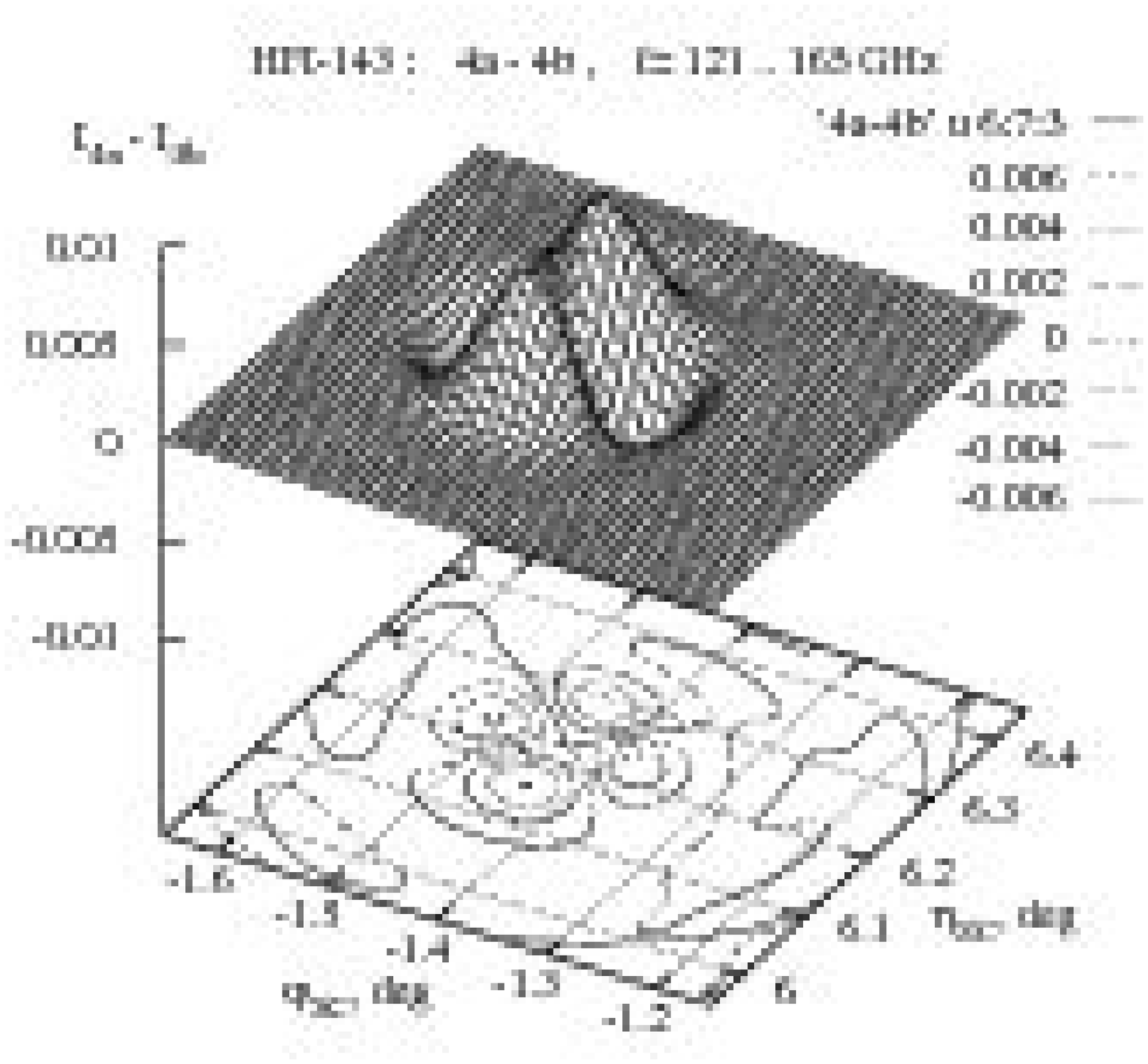}
(a)
   \includegraphics[width=8cm]{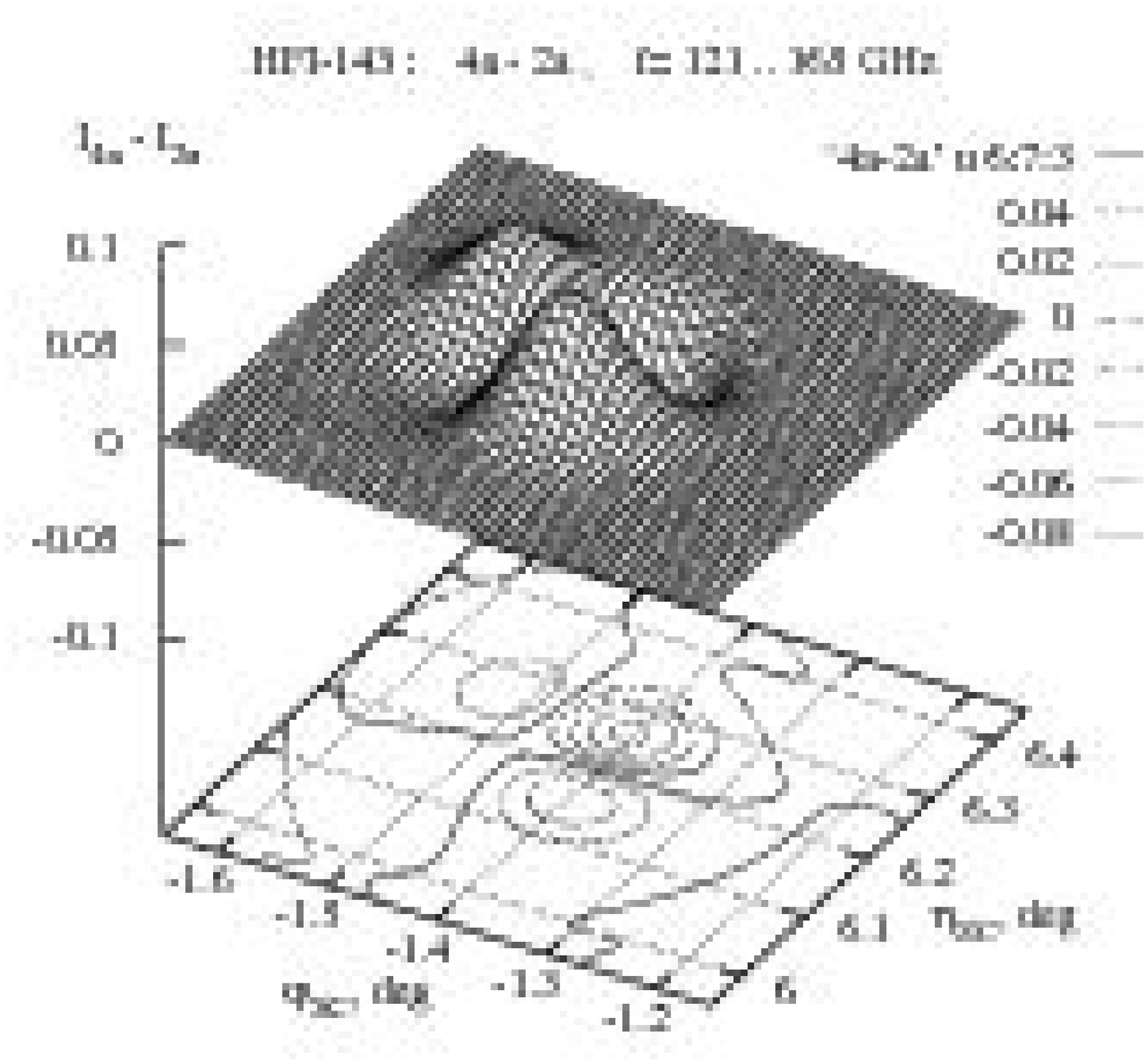}
(b)
      \caption{Relative difference of power patterns
(a) $\widetilde I_{4a} - \widetilde I_{4b}$ of two orthogonal channels 
of the same beam HFI-143-4 ($\psi=0^{\circ}$ and $\psi=90^{\circ}$, 
respectively) and (b) $\widetilde I_{4a} - \widetilde I_{2a}$ of the 
channels HFI-143-4a and HFI-143-2a ($\psi=0^{\circ}$ and 
$\psi=45^{\circ}$, respectively) when superimposed by spinning 
the telescope about the geometrical spin axis.
              }
         \label{beam_difference}
   \end{figure}

   \begin{figure}[tb]
   \centering
   \includegraphics[width=8cm]{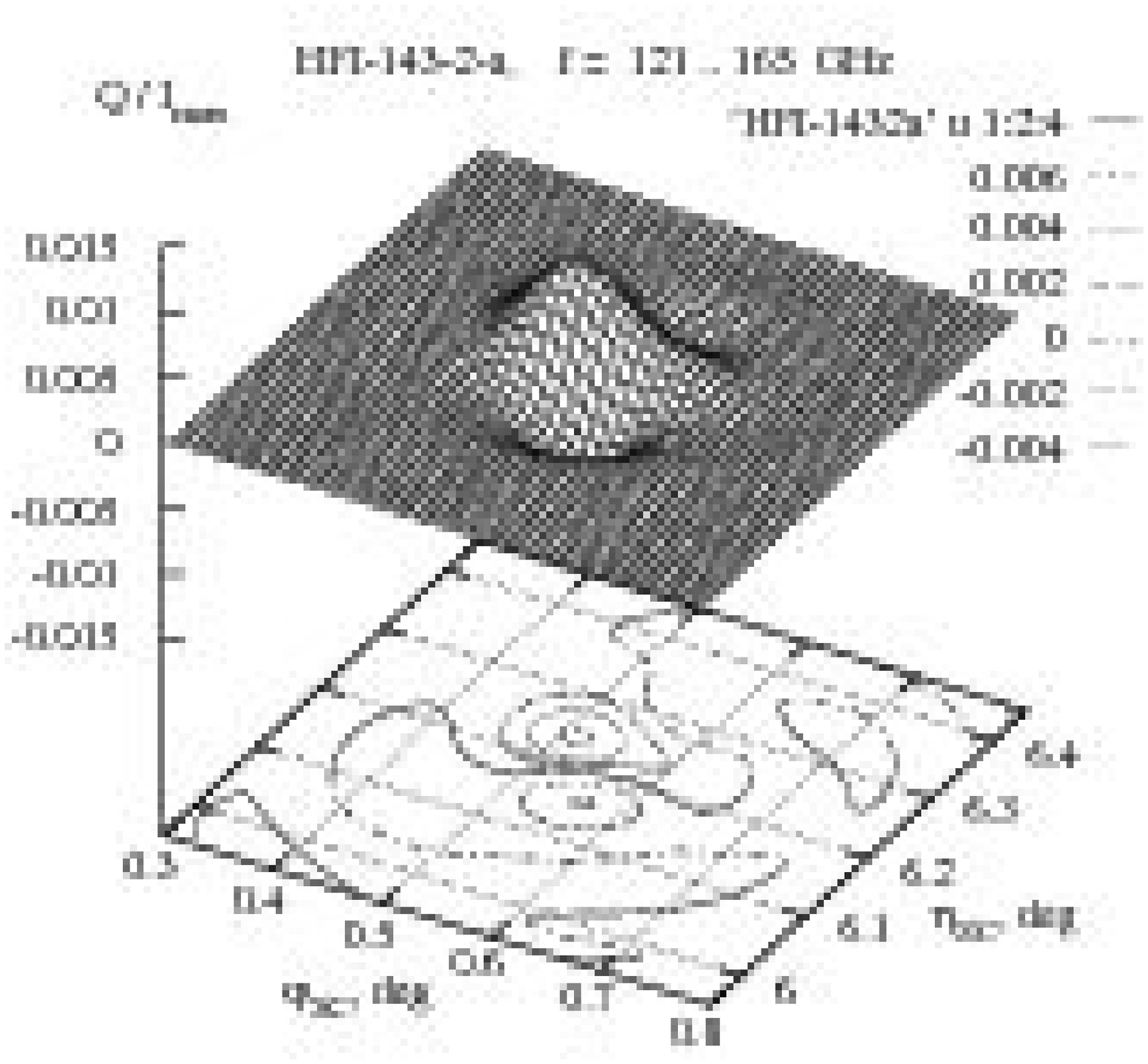}
(a)
   \includegraphics[width=8cm]{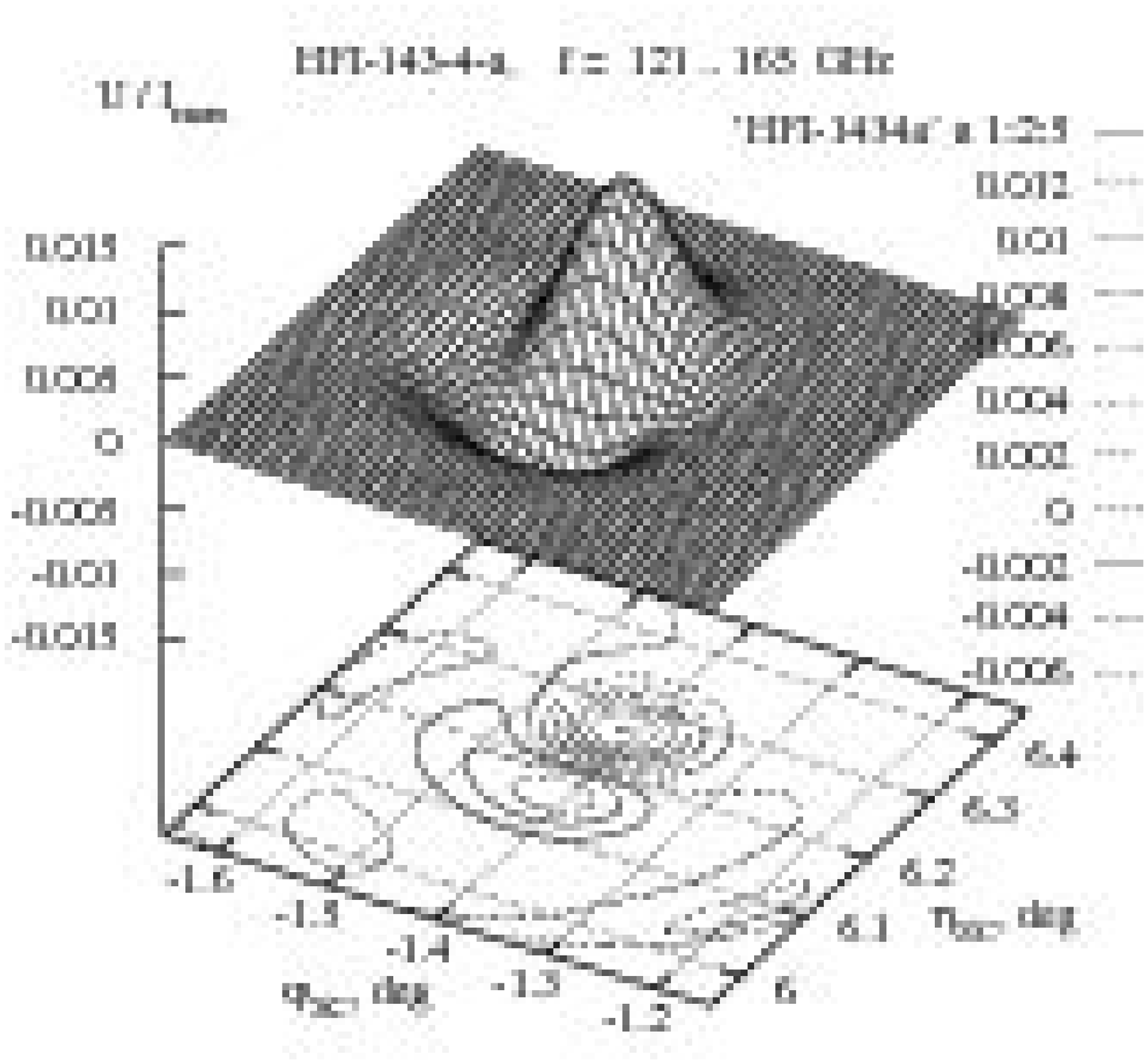}
(b)
      \caption{The $\widetilde Q$ and $\widetilde U$ Stokes parameters 
responses: (a) $\widetilde Q$ of the beam HFI-143-2a and (b) $\widetilde U$ 
of the beam HFI-143-4a (ideally, both $\widetilde Q$ and $\widetilde U$ 
should be zero in the beams of these polarizations for the selected 
reference axes).
              }
         \label{Stokes_parameters}
   \end{figure}

To minimize the beam mismatch between the channels of orthogonal
polarizations, the PSB bolometers of each pair of channels, $a$ and
$b$, are placed in the same cavity in the rear side of the respective
horn and share the same optics --- waveguides, filters, horns,
telescope. Because of this design, the difference of power patterns of
orthogonal polarizations of the same beam on the sky is really small
(Fig.~\ref{beam_difference}, a), with the peak value for all the
broadband beams being about $\delta \widetilde I_{4a4b} ={\rm
max}(\widetilde I_{4a}-\widetilde I_{4b})/\widetilde I_{max}=0.6\%$
(at the central frequency $f=143$ GHz, the typical difference is
$\delta \widetilde I_{4a4b}=0.9\%$).

A small difference of this kind arises for two reasons: (a) due to minor 
axial asymmetry of polarized modes that appears on the horn aperture 
when the PSB radiation (in the transmitting mode) propagates through 
the horn (the difference varies in sign and magnitude with frequency, 
though being well balanced over the band) and (b) due to some difference 
in the propagation of different polarizations along the same path via the 
telescope (all the differences are computed with the patterns normalized 
to the unit total power of the beams).

The mismatch of power patterns of different beams is about 10 times
more significant (Fig.~\ref{beam_difference}, b).  It depends
essentially on the location of horns in the focal plane of telescope.
For the pair of beams HFI-143-2 and HFI-143-4, when superimposed on
the sky by spinning the telescope until the coincidence of azimuths of
beam axes, the peak difference of the relative power across the
pattern varies from $7.0\%$ to $8.2\%$ depending on the polarizations
being compared.  Notice that the statistical difference of $5\%$ is
already rather crucial for the reliable reconstruction of the CMB
polarization map (Kaplan et al, \cite{Kaplan}).

Fig.~\ref{Stokes_parameters} shows the patterns of $\widetilde Q$ and
$\widetilde U$ Stokes parameter responses of the HFI-143-2a and
HFI-143-4a beams, respectively.  The peak values of these parameters
are $\widetilde Q_{2a}=0.6\%$ and $\widetilde U_{4a}=1.2\%$ (ideally,
$\widetilde Q$ and $\widetilde U$ should be zero in these
polarizations). For comparison, the peak values of $\widetilde V$ are
$3.7\%$ and $4.2\%$, respectively. The positive and negative values of
$\widetilde Q$ and $\widetilde U$ (as well as $\widetilde V$) are well
balanced over the beam patterns and the average is very close to
zero. It proves that the chosen directions of polarization of the horn
aperture field as found by optimizing on-axis beam polarization
directions (Yurchenko,\cite{Cervinia}) are pretty good, even though
the beam patterns are not quite symmetrical due to aberrations.

The analysis of different contributions to non-zero values of
$\widetilde Q_{2a}$, $\widetilde U_{4a}$ and $\widetilde V$ shows that
$\widetilde V$ arises mainly because of the field propagation via the
telescope ($\widetilde V_{\rm max}$ on the horn aperture is only
$0.5\%$).  On the contrary, both $\widetilde Q_{2a}=0.6\%$ and
$\widetilde U_{4a}=1.2\%$ given above and the power differences
between the orthogonal channels of the same beams, $\delta\widetilde
I_{2a2b}=\delta \widetilde I_{4a4b} =0.6\%$, are essentially due to
the horn effects ($\widetilde U =0.4\%$ and $\delta \widetilde
I_{ab}=0.8\%$ on the horn aperture). The telescope contribution,
though non-additive, is still important, as the propagation of the
axially symmetric quasi-Gaussian source field shows (in this case,
$\delta \widetilde I_{2a2b}=0.9\%$ and $\delta \widetilde
I_{4a4b}=1.0\%$ in the beams on the sky, being zero in the source
field).

Finally, in the cross-beam power differences $\delta \widetilde
I_{2\alpha 4\beta}$, both the horn and the telescope effects are
significant (e.g., $\delta \widetilde I_{2\alpha 4\beta}$ depends, to
some extent, on polarizations being compared), although the telescope
effect dominates (for the quasi-Gaussian source field, $\delta
\widetilde I_{2\alpha 4\beta}$ varies from $5.7\%$ to $6.7\%$ in a way
consistent with the variations in the beams from the actual corrugated
horns of respective polarizations).


\end{document}